\documentclass{revtex4-2}
\usepackage{mathtools}
\usepackage{natbib}
\usepackage{graphicx}
\usepackage{amsmath}
\usepackage{amssymb}
\usepackage{bm}
\usepackage{url}
\usepackage{hyperref}
\usepackage{cleveref}
\usepackage{booktabs}
\usepackage[flushleft]{threeparttable}
\usepackage{subcaption}
\usepackage{soul}
\usepackage{orcidlink}
\usepackage{comment}

\makeatletter
\let\jnl@style=\rm
\def\ref@jnl#1{{\jnl@style#1}}
\def\apjl{\ref@jnl{ApJ}}                
\def\aap{\ref@jnl{A\&A}}                
\def\mnras{\ref@jnl{MNRAS}}             
\def\aaps{\ref@jnl{A\&AS}}              
\makeatother

\begin{document}

\title{Can the symmetric \textit{Fermi} and eROSITA bubbles be produced by tilted jets?}

\author{\href{https://orcid.org/0000-0002-1868-0660}{Po-Hsun Tseng}}
\affiliation{Institute of Astrophysics, National Taiwan University, Taipei 10617, Taiwan}
\affiliation{Department of Physics, National Taiwan University, Taipei 10617, Taiwan}

\author{\href{https://orcid.org/0000-0003-3269-4660}{H.-Y. Karen Yang}}
\affiliation{Institute of Astronomy, National Tsing Hua University, Hsinchu 30013, Taiwan}
\affiliation{Department of Physics, National Tsing Hua University, Hsinchu 30013, Taiwan}
\affiliation{Physics Division, National Center for Theoretical Sciences, Taipei 10617, Taiwan}

\author{\href{https://orcid.org/0009-0003-2783-6836}{Chun-Yen Chen}}
\affiliation{Institute of Astrophysics, National Taiwan University, Taipei 10617, Taiwan}

\author{\href{https://orcid.org/0000-0002-1249-279X}{Hsi-Yu Schive}}
\affiliation{Institute of Astrophysics, National Taiwan University, Taipei 10617, Taiwan}
\affiliation{Physics Division, National Center for Theoretical Sciences, Taipei 10617, Taiwan}
\affiliation{Department of Physics, National Taiwan University, Taipei 10617, Taiwan}
\affiliation{Center for Theoretical Physics, National Taiwan University, Taipei 10617, Taiwan}

\author{\href{https://orcid.org/0000-0003-2654-8763}{Tzihong Chiueh}}
\affiliation{Institute of Astrophysics, National Taiwan University, Taipei 10617, Taiwan}
\affiliation{Physics Division, National Center for Theoretical Sciences, Taipei 10617, Taiwan}
\affiliation{Department of Physics, National Taiwan University, Taipei 10617, Taiwan}

\begin{abstract}
The \textit{Fermi Gamma-Ray Space Telescope} reveals two large bubbles in the Galaxy,\
extending nearly symmetrically $\sim50^{\circ}$ above and below the Galactic center (GC).\
Previous simulations of bubble formation invoking active galactic nucleus (AGN) jets\
have assumed that the jets are vertical to the Galactic disk;\
however, in general,
the jet orientation does not necessarily correlate with the rotational axis of the Galactic disk.
Using three-dimensional special relativistic hydrodynamic simulations\
including cosmic rays (CRs) and thermal gas, we show that the dense clumpy gas within the Galactic disk\
disrupts jet collimation  (``failed jets" hereafter),
which causes the failed jets to form
hot bubbles. Subsequent buoyancy in the stratified atmosphere renders them vertical to form
the symmetric \textit{Fermi} and eROSITA bubbles (collectively, Galactic bubbles).
We find that\
(1) despite the relativistic jets emanated from the GC\
are at various angles
$\leq 45^{\circ}$ with respect to
the rotational axis of the Galaxy, the Galactic bubbles nonetheless appear aligned with the axis;
(2) the edge of the eROSITA bubbles corresponds to a forward shock driven by the hot bubbles;
(3) followed by the forward shock is a tangling contact discontinuity corresponding to the edge of the \textit{Fermi} bubbles;
(4) assuming a leptonic model we find that the observed gamma-ray bubbles and microwave haze can be reproduced with
a best-fit CR power-law spectral index of 2.4;
The agreements between the simulated and the
observed multi-wavelength features suggest that forming the Galactic bubbles by
oblique AGN failed jets is a plausible scenario.
\end{abstract}


\maketitle

\section{Introduction} \label{sec:intro}

The detection of the \textit{Fermi} bubbles \citep{Su2012,Ackermann2014,Narayanan2017},\
two large bubbles symmetrically extending about 50 degrees above and below the Galactic plane,\
is one of the great discoveries of the \textit{Fermi} Large Area Telescope \citep{Atwood2009}.\
The gamma-ray emission of the \textit{Fermi} bubbles is observed in the energy range\
of $\sim$1--100 GeV and has an almost spatially uniform hard\
spectrum, sharp edges, and an approximately flat brightness distribution (see reviews by  \citealt{Yang2018} and \citealt{Yang2023}). Recently, the newly launched eROSITA \citep{Predehl2021} conducted an all-sky X-ray survey with high sensitivities and revealed two gigantic bubbles (eROSITA bubbles hereafter) extending to $\sim 80$ degrees in Galactic latitudes, corresponding to an intrinsic size of 14 kpc across \citep{Predehl2020}.\
The remarkable resemblance between the eROSITA and \textit{Fermi} bubbles suggests that they likely share the same origin \citep{Yang2022}.
Their symmetry about the GC further suggests that these Galactic bubbles may be generated by powerful energy injections from the GC, possibly related to nuclear star formation\
\citep{PhysRevLett.106.101102,Carretti2013,Crocker2015,Sarkar2015}
or past AGN activity \citep{Guo2012,Guo2012b,Yang2012,Yang2013,Mou2014,Yang2017}. The latter scenario is what we will focus on in this work.

Previous attempts \citep{Guo2012,Yang2012,Zhang2020}\
to model the formation of the symmetric Galactic bubbles by AGN jets have typically\
assumed that the jets are vertical to the Galactic plane. While there are some observational indications of pc-scale jets from Sgr A* that are found to be perpendicular to the Galactic plane \citep{Li, Zhu}, generally speaking, the AGN jet orientation is determined by the black hole spin and the accretion disk in the black-hole vicinity and does not need to align with the rotational axis of the host galaxy. Indeed, observationally there is a lack of evidence for the alignment between AGN jets and the disk normal \citep[e.g.,][]{Gallimore2006}. The jets are often oblique to the disk normal\
(e.g. NGC 3079, \citealt{Cecil2001}; NGC 1052, \citealt{Dopita2015}),\
and there are even cases in which the jets lie in the plane of the disk (e.g., IC 5063, \citealt{Morganti2015}).\

To this end, the aim of this work is to remove the assumption on jet orientations in the AGN jet models by introducing a dense, thin interstellar medium (ISM) disk\
that can interact with the central oblique jet,\
in an attempt to resolve the symmetry problem of the Galactic bubbles. More specifically, we use three-dimensional\
special relativistic hydrodynamics (SRHD) simulations\
involving CR jet injections from the central supermassive black hole (SMBH) in the Galaxy to investigate\
whether the \textit{oblique} jet scenario is able to produce\
the \textit{symmetric} Galactic bubbles. We will verify whether the oblique jet model is\
consistent with the observed features of the Galactic bubbles, including the shape, \
surface brightness, and spectra of the \textit{Fermi} bubbles \citep{Ackermann2014}\
and microwave haze \citep{Dobler_2008,PlanckCollaborationIX2013}.

This paper is organized as follows.\
In Section \ref{Methodology}, we describe the numerical techniques and initial conditions employed.\
In Section \ref{Results}, we first present characteristics of our simulated Galactic bubbles,\
and then discuss how the disk affects the formation of the bubbles.\
We compare the morphology and profiles of the simulated eROSITA bubbles with\
the observed X-ray map in Section \ref{X-ray},\
and present the simulated and observed multi-wavelength spectra of\
the \textit{Fermi} bubbles 
in Section \ref{sec:gamma-ray-microwave}.\
Finally, the discussion and conclusion of our findings are given in Section \ref{sec:discussion} and  \ref{Conclusions}, respectively.

\section{Methodology}
\label{Methodology}
  We use the GPU-accelerated SRHD adaptive-mesh-refinement (AMR) code \textsc{gamer-sr} developed at the\
  National Taiwan University\
  (\citeauthor{gamer-1} \citeyear{gamer-1}, \citeyear{gamer-2}; \citeauthor{tseng2021} \citeyear{tseng2021})\
  to carry out the simulations of the Galactic bubbles formed by CR and relativistic-fluid injections from the GC.

  The governing equations solving the special relativistic ideal fluid\
  including CR advection, and dynamical coupling between the thermal gas and CRs without CR diffusion\
  can be written in a succinct form as

  \begin{subequations}
    \label{governing-eq}
    \begin{align}
     &\partial_{t} D+\partial_{j} \left(DU^{j}/\gamma\right)=0,\label{D evolution}\\
     &\partial_{t} M^{i}+\partial_{j} \left(M^{i}U^{j}/\gamma+p_{\text{total}}\delta^{ij}\right)=\
     -\rho\partial_{i}\Phi,\label{M evolution}\\
     &\partial_{t} \tilde{E}+\partial_j \left[\left(\tilde{E}+p_{\text{total}}\right)U^{j}/\gamma\right]=-\rho(U_{j}/\gamma)\partial_{j}\Phi, \label{E evolution}\\
     &\partial_{t} \left(\gamma e_{\text{cr}}\right) + \partial_{j} \left(e_{\text{cr}}U^{j}\right)=\
     -p_{\text{cr}} \partial_{j} U^{j}-p_{\text{cr}}\partial_{t}\gamma, \label{Ecr evolution}
    \end{align}
  \end{subequations}

  where the five conserved quantities of gas $D$, $M^{i}$, and $\tilde{E}$ are the mass density,\
  the momentum densities, and the reduced energy density, respectively.\
  The reduced energy density is defined by subtracting the rest mass energy density of gas\
  from the total energy density of gas.\
  $\gamma$ and $U^{j}$ are the temporal and spatial components of four-velocity of gas.\
  $\rho$ is the gas density in the local rest frame defined by $D/\gamma$.\
  $p_{\text{gas}}$ is the gas pressure.\
  $p_{\text{cr}}$ and $e_{\text{cr}}$ are the CR pressure and CR energy density measured in the local rest frame, related by $p_{\text{cr}}=e_{\text{cr}}/3$.\
  $p_{\text{total}}$ is the sum of $p_{\text{gas}}$ and $p_{\text{cr}}$.\
  $\Phi$ is the gravitational potential, $G$ is the gravitational constant,
  $c$ is the speed of light, and $\delta^{ij}$ is the Kronecker delta notation.\
  Throughout this paper, Latin indices run from 1 to 3, except when stated otherwise. The set of Eq. \ref{governing-eq} is closed by using the Taub-Mathews equation of state \citep[EoS;][]{Taub,TM_EOS}\
  that approximates the exact EoS \citep{Synge} for ultra-relativistically\
  hot gases coexisting with non-relativistically cold gases. Note that (1) we simply replace $p_{\text{total}}$ by $p_{\text{gas}}$ in this paper\
  as we have assumed $p_{\text{cr}}\ll p_{\text{gas}}$; (2) we use the Newtonian gravity as an approximation since the gravitational effects resulting from the kinetic/thermal energies within jet sources are relatively insignificant compared to the overarching gravitational potential of the Galaxy; (3) the term $-p_{\text{cr}\partial_{t}}\gamma$ in Eq. \ref{Ecr evolution} is omitted from our implementation since the maximum fluid velocity across the simulation domain, $\gamma\sim1.166$ --- found at the jet source ---, is near the lower limit of the Lorentz factor,  rendering this term less significant in comparison to the term $-p_{\text{cr}}\partial_{j}U^{j}$.

  \textsc{gamer-sr} adopts a new SRHD solver \citep{tseng2021}, which significantly reduces numerical errors
  in non- and ultra-relativistic limits caused by catastrophic cancellations
  in the conversion between primitive ($\rho$, $U^{j}$, $p$) and conserved variables ($D$, $M^{j}$, $\tilde{E}$).
  \textsc{gamer-sr} also adaptively and locally reduce the min-mod coefficient\
  \citep{tseng2021} within the failed patch group rarely occurring in the SRHD solver,\
  new patches allocations, and ghost-zone interpolations.\
  In this manner, we provide an elegant approach to avoid the use of pressure/density floor,\
  which is unnatural but widely used in almost all publicly available codes.\

  In order to track the evolution of CRs injected by the AGN jets and make predictions of the non-thermal radiation they produce, we adopt the CR hydrodynamic formalism and model the CRs as a second fluid \citep{Zweibel2013}. The approach is similar to previous works of \cite{Guo2012} and \cite{Yang2012}, but generalized to CRs that couple with thermal gas moving with relativistic speeds. 
  The detailed implementations of CR in \textsc{gamer-sr} and tests of algorithm can be found in Appendix \ref{appendix_B}.
  In this approach, the CRs are treated as a single species without distinctions between CR electrons and protons,\
  and the CR energy density $e_{\text{cr}}$ is evolved according to Eq. \ref{Ecr evolution}.
  The CRs are advected with the thermal gas and can have adiabatic compression and expansion with the gas. Also, we do not simulate the spectral evolution of the CRs and assume that the CR-to-gas pressure ratio is much less than 1 so that\
  the contribution of CR pressure gradient to the momentum of the gas can be ignored\
  (we will see that the ratio is around 0.005--0.15 throughout the simulations).
Therefore, in the simulations we have neglected the cooling of CRs because it should have a negligible impact on the overall dynamics.

  As stressed by \citet{Yang2012}, CR diffusion with a canonical diffusion coefficient of $\kappa \sim 3\times 10^{28}$ cm$^2$ s$^{-1}$ in the Galaxy has a minor effect\
  on the overall morphology of the \textit{Fermi} bubbles as it
  only acts to smooth the CR distributions on the scales of $l \sim \sqrt{\kappa t} \sim 0.3\ (t/1 {\rm Myr})$ kpc. Including anisotropic CR diffusion can also help to sharpen the edges of the bubbles due to interplay between the magnetic field\
  and anisotropic CR diffusion with suppressed perpendicular diffusion across the bubble surface. As for the magnetic field, \cite{Yang2013} has found that the magnetic field within the \textit{Fermi} bubbles needs to be amplified to comparable values to the ambient field in order to reproduce the microwave haze emission. We thus directly adopt the exponential model for the magnetic field distribution in our calculation for the haze (see descriptions in Section \ref{sec:model}).
   For the above reasons, we have ignored CR diffusion and the magnetic field in the simulations.\

 In our study, the simulation domain is\
  $14\times14\times28$ kpc$^3$ in size, slightly larger than the observed eROISTA bubbles, with the outflow boundary condition. The domain is resolved using the Cartesian coordinate system with the finest spatial resolution 0.4 pc, approximately 50 times smaller than a typical molecular cloud.

  \subsection{The Galactic halo and disk models} \label{sec:model}
  As a proof-of-concept study, we approximate conventionally axisymmetric stellar potential of Milky Way\
  by a plane-parallel potential that is symmetric about the Galactic plane, $z=0$,\
  in the simulation domain. The plane-parallel potential is fixed throughout our simulations and given by
  \begin{equation}
    \label{total-gravitational-potential}
    \Phi_{\text{total}}(z) = \Phi_{\text{bulge}}(z) + \Phi_{\text{halo}}(z),
  \end{equation}
  where
  \begin{equation}
    \Phi_{\text{bulge}}(z)=\
    2\sigma^2_{\text{bulge}}\
    \ln\cosh\left(z\sqrt{\frac{2\pi G\rho_{\text{bulge}}^{\text{peak}}}{\sigma^2_{\text{bulge}}}}\right)
  \end{equation}
  is the potential of an isothermal slab mainly contributed by stars around the Galactic bulge, and\
  $\Phi_{\text{halo}}(z)=v^2_{\text{halo}}\ln\left(z^2+d^2_{\text{h}}\right)$\
  is a plane-parallel logarithmic dark halo potential. 
  The plane-parallel geometry for the Galactic halo is a simplified assumption that allows us to focus on the interaction between the jets and the clump disk (Section \ref{sec:clump_disk}). In reality, the Galactic halo distribution should also depend on the galactocentric radius, as adopted in previous jet simulations \citep[e.g.,][]{Guo2012, Sarkar2015, Yang2022, Sarkar_2023}. Thus, the choice of plane-parallel geometry balances reality and the specific scientific questions we seek to answer. The radial potential profile and the centrifugal forces introduced by the 
  rotating Galaxy may somewhat modify the morphology of the bubbles wider than the Galactic bulge after the jets break out from the dense disk; modified jet parameters may be needed to match with the morphology of the observed bubbles. One should therefore be cautious when making comparisons with the previous simulations.

  With the isothermal assumption and the condition of hydrostatic equilibrium within the total potential of the disk and halo, as well as pressure equilibrium between the isothermal disk and the halo gas, we can write down the steady-state gaseous density distribution as\
  \begin{subequations}
  \begin{align}
     \displaystyle \rho_{\text{isoDisk}}(z) = \rho_{\text{isoDisk}}^{\text{peak}}
     \exp\left[-\frac{\Phi_{\text{total}}(z)}{k_{B}T_{\text{isoDisk}}/m_{\text{p}}}\right]&\label{isothermal-disc-density}\\
     \text{, if $|z| < z_{0}$,}& \nonumber \\
     \nonumber\\
     \displaystyle \rho_{\text{atmp}}(z) = \rho_{\text{atmp}}^{\text{peak}}
     \exp\left[-\frac{\Phi_{\text{total}}(z)}{k_{B}T_{\text{atmp}}/m_{\text{p}}}\right]&\label{isothermal-atmp-density}\\
     \text{, otherwise,}& \nonumber
  \end{align}
  \label{disc-atm-sys}
  \end{subequations}
  where $m_{\text{p}}$ is the proton mass,\
  $T_{\text{isoDisk}}$ and $T_{\text{atmp}}$ are the temperatures of the isothermal disk and the ambient atmosphere, and\
  $\rho_{\text{isoDisk}}^{\text{peak}}$ and $\rho_{\text{atmp}}^{\text{peak}}$ are the peak density\
  of the disk and the atmosphere at $z=0$, respectively.

\begin{figure}
    \includegraphics[width=\columnwidth]{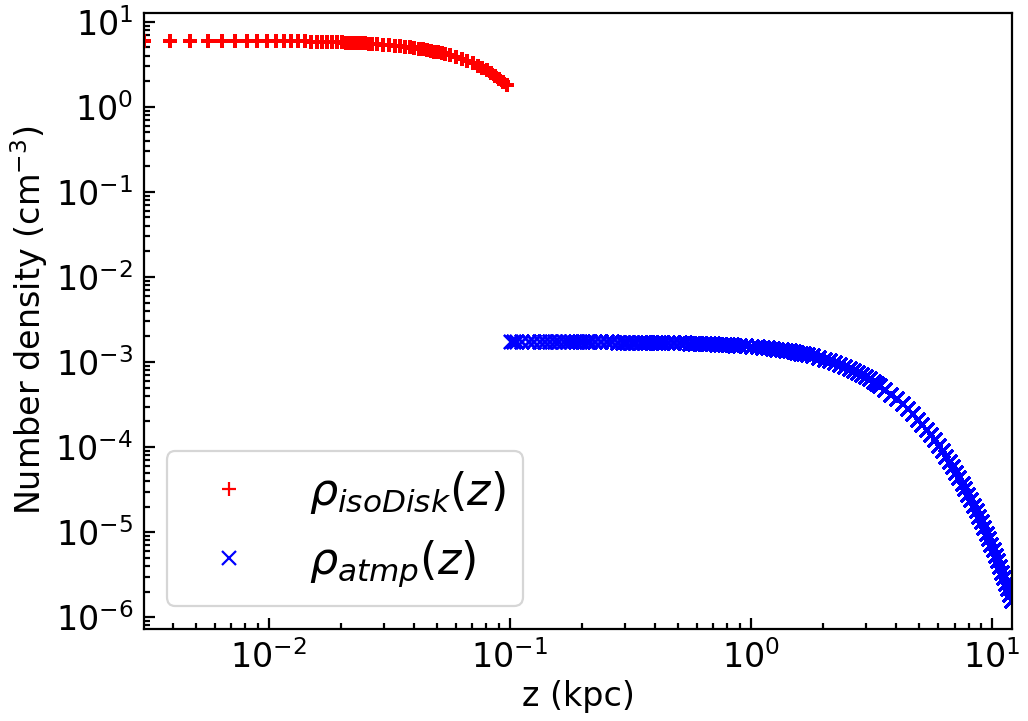}
    \caption{The density profile of the isothermal disk (red pluses) and
             the ambient atmosphere (blue crosses) along the positive z-axis.
             The density distribution is derived from the condition of hydrostatic equilibrium.
             The gas at the interface between the isothermal disk
             and the atmosphere at
             $z=0.1$
             is in pressure
             equilibrium.}
    \label{fig__initial-density-profile}
  \end{figure}

  We tabulate parameters adopted for the Galactic model in \Cref{table-parameters},\
  except for $\rho_{\text{atmp}}^{\text{peak}}$ that\
  can be derived from the other known parameters and the pressure equilibrium condition\
  at the interfaces $(z=\pm z_{0})$ between the disk and the atmosphere. The density profile of Eq. \ref{disc-atm-sys} is shown in Fig. \ref{fig__initial-density-profile}.\
  Beyond the core radius ($\sim 2 \text{ kpc}$) the gas density decreases rapidly as a power-law.

  To compute the predicted synchrotron radiation as a function of position,\
  we adopt the default exponential magnetic field in \textsc{galprop} \citep{Strong2007}
  that obeys the following spatial dependence:\

  \begin{equation}
     |\mathbf{B}(R, z)|=B_{0}\exp\left[-\frac{z}{z_{0}}\right]\exp\left[-\frac{R}{R_{0}}\right],
     \label{magnetic-field}
  \end{equation}

  where $R=\sqrt{x^{2}+y^{2}}$, $B_{0}$ is the average field strength at the GC,\
  and $z_{0}$ and $R_{0}$ are the characteristic scales in the vertical and radial\
  directions, respectively. We adopt $z_{0} = 2$ kpc and $R_{0} = 10$ kpc, which\
  are best-fitting values in the \textsc{galprop} model to reproduce the\
  observed large-scale 408 MHz synchrotron radiation in the Galaxy.\
  We choose $B_{0} = 50$ $\mu$G based on the observed field strength at the\
  GC \citep{Crocker2010}.


  \subsection{The clumpy multiphase interstellar medium}
  \label{sec:clump_disk}

  A crucial component in our work is the clumpy ISM disk initialized by\
  the publicly available pyFC code\footnote{\url{https://pypi.python.org/pypi/pyFC}}.
  pyFC randomly generates a dimensionless 3D scalar field $f(\bold{x})$\ 
  that obeys the log-normal probability distribution\
  with mean $\mu$ and dispersion $\sigma$,\
  and follows the power-law Kolmogorov spectrum
  \begin{equation}
    D(\bold{k})=\int k^{2} \hat{f}(\bold{k})\hat{f}^{*}(\bold{k})d\Omega \propto k^{-\delta},
    \label{Kolmogorov-spectrum}
  \end{equation}
  where $\hat{f}(\bold{k})$ is the Fourier transform of $f(\bold{x})$.\
  The spectrum $D(\bold{k})$ in the Fourier space is characterized by a power-law index $\delta=5/3$,\
  a Nyquist limit $k_{\text{max}}$, and a lower cutoff wave number $k_{\text{min}}$.
  $k_{\text{max}}$ is one-half of the spatial resolution within the disk,\
  and $k_{\text{min}}$ is 375.0, corresponding to the maximum size of an individual clump of $\sim 20$ pc.\
  \citet{LA2002} and \citet{Wagner2012} have outlined a detailed procedure\
  for constructing a clumpy scalar field, and we refer the readers for more information.

  The density of the clumpy disk can then be obtained by taking the scalar products of\
  $f(\bold{x})$ with $\rho_{\text{isoDisk}}(z)$ over all cells within the disk, i.e.,\
  $\displaystyle\rho_{\text{ismDisk}}(\bold{x}) =\
  f(\bold{x}) \rho_{\text{isoDisk}}(z)$.\
  Also, the thermal pressure equilibrium within the clumpy disk implies that the temperature of the disk is
  $\displaystyle T_{\text{ismDisk}}(\bold{x}) =\
  T_{\text{isoDisk}}(z)\rho_{\text{isoDisk}}(z)/\rho_{\text{ismDisk}}(\bold{x})$. The last category in \Cref{table-parameters} summarizes the parameters of the clumpy disk and their references.

\begin{figure}
      \includegraphics[width=\columnwidth]{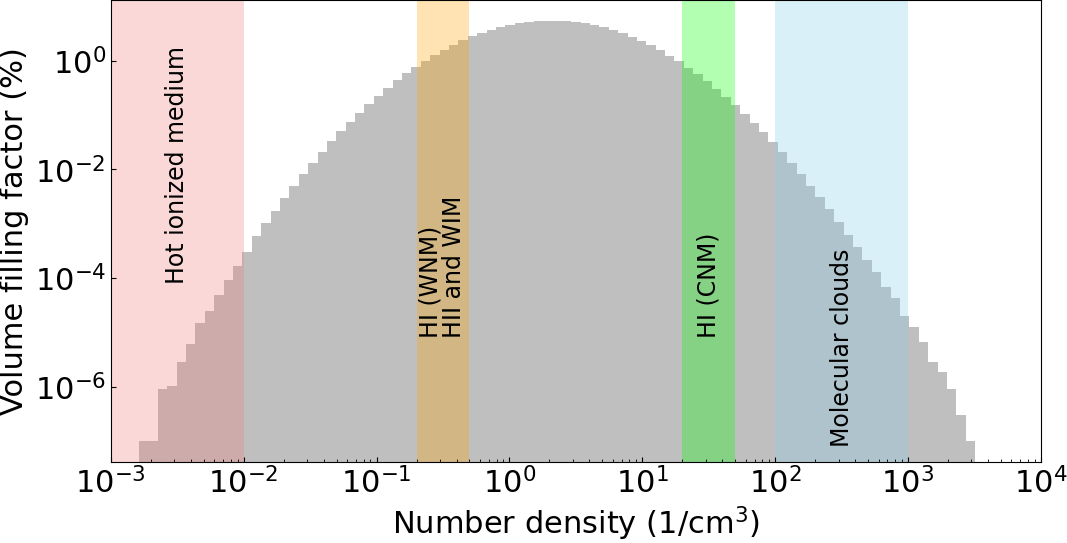}
    \caption{The volume filling factor as a function of\
             initial number density within the disk without the jet source.\
             The vertical bands from left to right depict the allowable number densities \citep{peak-ism-density} for\
             hot ionized, warm neutral (WNM), warm ionized (WIM), cold neutral mediums (CNM), and molecular clouds.}
      \label{fig__numberDensityHistogram}
  \end{figure}

\begin{figure}
    \includegraphics[width=\columnwidth]{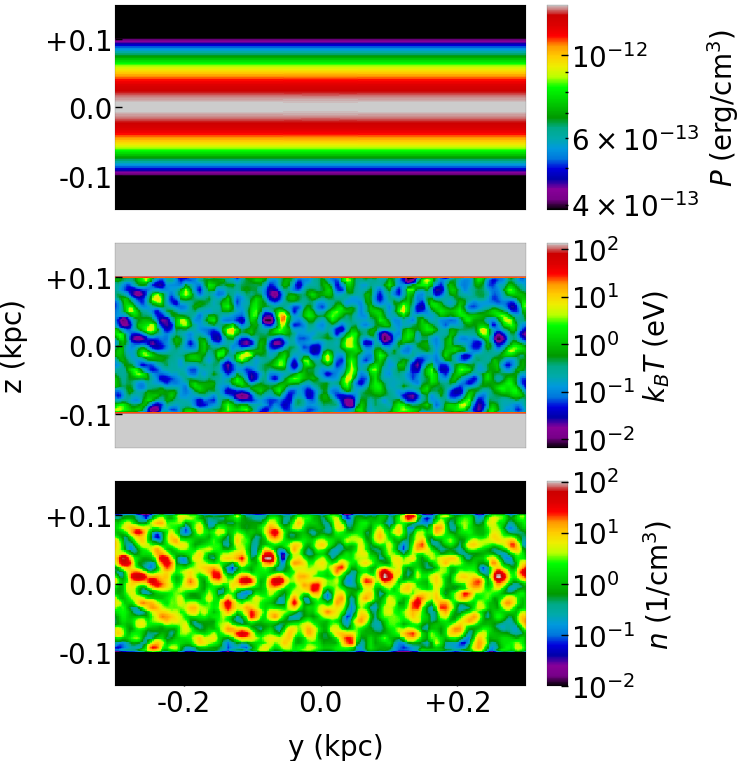}
    \caption{Close-up view of the initial\
             pressure (top), temperature (middle), and number density (bottom) slices\
             in the $y-z$ plane through the center of the disk.
             }
    \label{fig__zoom-in-disc}
  \end{figure}

  On the basis of this setup, we cover the AMR base level with\
  $16\times16\times32$ root cells, refined progressively on the mid-plane at $z=0$\
  based on the gradient of density.\
  We also restrict the refinement level at 7 within the disk so that\
  a molecular cloud can be adequately resolved by approximately 30 cells along their diameter of 20 pc. We plot the volume filling factor as a function of\
  initial number density within the disk without the jet source in Fig. \ref{fig__numberDensityHistogram},\
  and show a close-up view of the\
  pressure, temperature, and number density slices
  in the $y-z$ plane through the center of the disk in Fig. \ref{fig__zoom-in-disc}.

\begin{table*}
\raggedright
\caption{Parameters of the disk, atmosphere, and gravitational potential in the simulations.}
\label{table-parameters}
\begin{tabular}{@{}llrc@{}}
\toprule[1pt]\midrule[0.3pt]
Parameter                             & Description                    & Value                                &  Reference                     \\ \midrule
{\bf Static stellar potential }       &                                &                                      &                                \\
$\sigma_{\text{bulge}}$               & Velocity dispersion of bulge   & 100 km s$^{-1}$                      & \citep{velocity-dispersion-MW} \\
$\rho_{\text{bulge}}^{\text{peak}}$   & Peak average density of bulge  & $4\times 10^{-24}$ g cm$^{-3}$       &   N/A                          \\ \hline
{\bf Static dark halo potential }     &                                &                                      &                                \\
$v_{\text{halo}}$                     & Characteristic velocity        & 131.5 km s$^{-1}$                    & \citep{Johnston1995}           \\
$d_{\text{h}}$                        & Core radius                    & 12 kpc                               & \multicolumn{1}{c}{''}         \\ \hline
{\bf Atmosphere }                     &                                &                                      &                                \\
$T_{\text{\text{atmp}}}$              & Temperature of atmosphere      & $10^{6}$ K                           & \citep{temperature-MW}         \\ \hline
{\bf Isothermal disk }                &                                &                                      &                                \\
$z_{0}$                               & Scale height of disk           & 100 pc                               & \citep{peak-ism-density}       \\
$T_{\text{\text{isoDisk}}}$           & Temperature of disk            & $10^{3}$ K                           & \multicolumn{1}{c}{''}         \\
$\rho_{\text{isoDisk}}^{\text{peak}}$ & Peak density of disk           & $10^{-23}$ g cm$^{-3}$               & \multicolumn{1}{c}{''}         \\ \hline
{\bf Clumpy disk }                    &                                &                                      &                                \\
$k_{\text{min}}^\dagger$              & Cutoff wave number             & 375.0                                & \citep{peak-ism-density}       \\
$\mu$                                 & Mean of scalar field           & 1.0                                  &   N/A                          \\
$\sigma^\ddag$                        & Dispersion of scalar field     & 5.0                                  & \citep{Federrath2010}          \\
$\delta$                              & Power law index                & -5/3                                 &   N/A                          \\ \midrule
\end{tabular}
\begin{tablenotes}
      \raggedright
      \item  $\dagger$  $k_{\text{min}}=375.0$ leads to the size of an individual molecular cloud of $\sim 100$ pc.
      \item  $\ddag$ In numerical simulations of turbulence,\
             \citet{Federrath2010} find $\sigma\sim 3.6$ and 35 for solenoidal (divergence-free)\
             and compressive (curl-free) driving force,\
             respectively, so that our adopted value of 5 is closer to their solenoidal result.
    \end{tablenotes}
\end{table*}

\subsection{Oblique jets}

  We simulate the jets emanating from the GC with an inclination angle\
  $45^{\circ}$ with respect to the Galactic plane\
  in order to alleviate the constraint that\
  the jet direction must be perpendicular to the Galactic plane, and in particular to\
  investigate how the dense disk affects bubble formation.

  We use the following quantities to characterize the jets:
  the flow four-velocity, $\beta\gamma=0.6$; the gas density, $\rho_{\text{jet}}=10^{-26}$ g/cm$^{3}$; the gas temperature, $k_{B}T=1.72$ MeV.\
  This temperature corresponds to a gas pressure ($p_{\text{gas}}$) of $1.65\times10^{-8}$ erg/cm$^3$. Additionally, the CR pressure ($p_{\text{CR}}$) is set to $7.0\times10^{-10}$ erg/cm$^{3}$ (i.e., the initial CR-to-gas pressure ratio $\sim$ $0.042$).\
  When compared to the surrounding medium, the density contrast between the thermal gas contained in the jet source and the ambient gas is\
  $\rho_{\text{jet}}/\rho_{\text{amb}}=10^{-3}$,\
  and the temperature contrast is $T_{\text{jet}}/T_{\text{amb}}=2\times10^{4}$.
  The jet power is thus $3.2\times 10^{42}$ erg s$^{-1}$, primarily driven by kinetic energy, with a kinetic-to-thermal energy ratio of 60.7 and an Eddington ratio of 0.008.\
  Note that since we inject the jets at the center of the clumpy disk,\
  we define the atmosphere gas density by the peak density of\
  the isothermal disk on the mid-plane $z=0$ (i.e. $\rho^{\text{peak}}_{\text{isoDisk}}$),\
  as opposed to the \textit{clumpy} density around the jet source.

  The bipolar jets originate from the center of the simulation domain (i.e., active zones). The gas properties inside the jets are reset for each evolution time step; thus, the jets are constantly ejected from a cylindrical source starting from the beginning of the simulation ($t=0$)\
  and suddenly quenched at $t=0.12$ Myr before fully breaking out the disk.\
  Without quenching, the Galactic bubbles at the present time would be asymmetric about the Galactic plane. The diameter and height of cylindrical source are 4 pc,\
  leading to a source volume ($\sim 50 \text{ pc}^{3}$)\
  much smaller than that of an individual clump by a factor of $\sim 83$.\
  By intentionally reducing the volume ratio of the jet source to an individual clump,\
  we can mitigate the effect of the randomness of the clumps on the bubbles.\
  Moreover, we resolve the jet source with the highest refinement level of 11,\
  bringing the finest spatial resolution up to 0.4 pc.\

  The duration of the jets permits a total ejected energy of $1.2\times10^{55}$ erg, which is 6--10 times lower than the estimated range of enclosed energy  falling between \
  $8\times10^{55}$ erg and $1.3\times10^{56}$ erg \citep{Predehl2020}. This discrepancy may arise from several assumptions involved in the observational constraint, such as the shape of the bubbles, thickness of the X-ray emitting shell, and the Mach number of the shock associated with the surface of eROSITA bubbles \citep{Predehl2020}.

\section{Results}
\label{Results}

\subsection{Morphology and properties of Galactic bubbles}

\begin{figure}%
      \captionsetup[subfigure]{labelformat=simple}
      \begin{subfigure}[b]{0.2\linewidth}%
         \includegraphics[height=10cm]{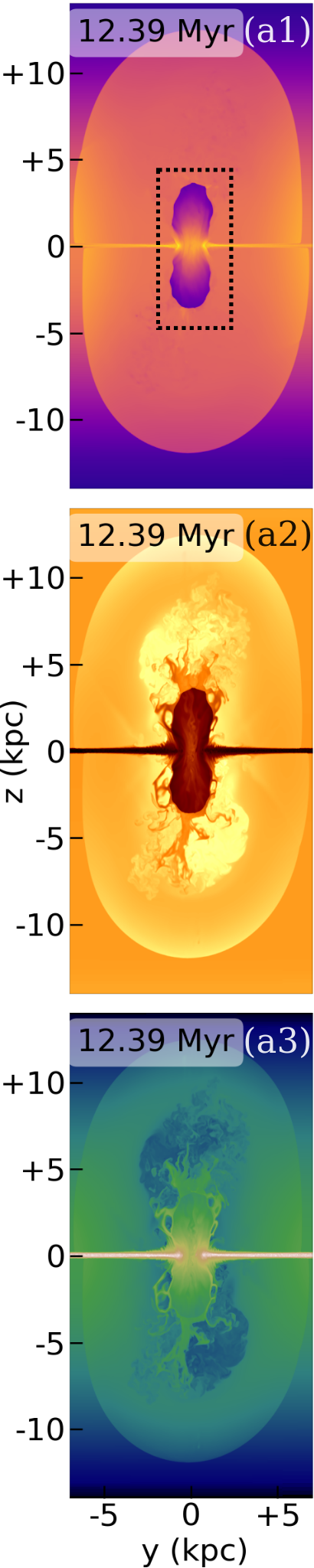}%
         \caption{}%
         \label{fig__jetI5+ismSeed3-45deg-a}%
      \end{subfigure}%
      \hspace{10pt}%
      \begin{subfigure}[b]{0.2\linewidth}%
         \includegraphics[height=10cm]{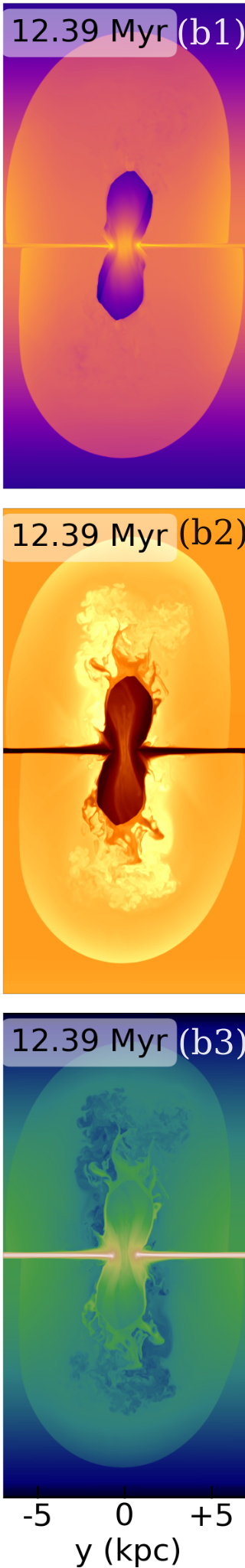}%
         \caption{}%
         \label{fig__jetI5+ismSeed3-45deg-b}%
      \end{subfigure}%
      \hspace{-2pt}%
      \begin{subfigure}[b]{0.2\linewidth}%
         \includegraphics[height=10cm]{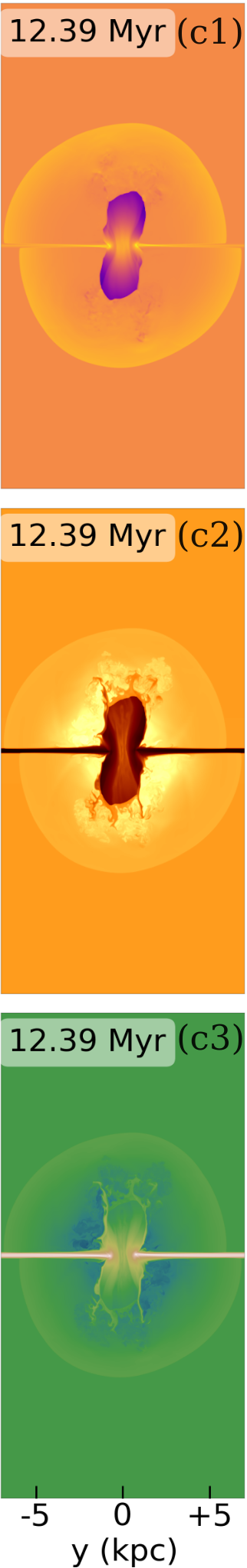}%
         \caption{}%
         \label{fig__jetI5+ismSeed3-45deg-c}%
      \end{subfigure}%
      \hspace{-2pt}%
      \begin{subfigure}[b]{0.2\linewidth}%
         \includegraphics[height=10cm]{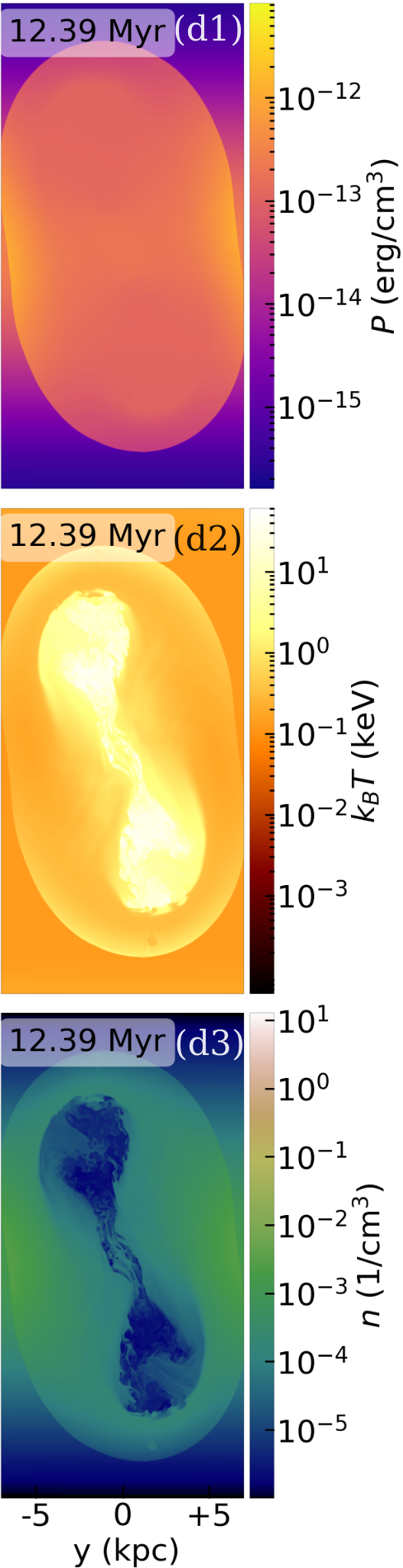}%
         \caption{}%
         \label{fig__jetI5+ismSeed3-45deg-d}%
     \end{subfigure}%
      \caption{
             The slices of pressure (top), temperature (middle), and number density (bottom)\
             at the end of simulation $t=12.39$ Myr.\
             The slices pass through a bipolar jet source injecting along $z=-y$ direction\
             for a duration $t=0$--$0.12$ Myr.\
             Comparison between the clumpy (Fig. \ref{fig__jetI5+ismSeed3-45deg-a})\
             and the smooth disks (Fig. \ref{fig__jetI5+ismSeed3-45deg-b}) in a stratified atmosphere\
             shows that the initial density distribution of the dense disk has an insignificant effect\
             on the overall dynamics of bubbles. However, the outermost bubbles arising from the smooth disk\
             in a uniform atmosphere (Fig. \ref{fig__jetI5+ismSeed3-45deg-c}) are nearly spherical,\
             suggesting that the stratification facilitates the elongation of the outermost bubbles significantly.\
             Fig. \ref{fig__jetI5+ismSeed3-45deg-c} and \ref{fig__jetI5+ismSeed3-45deg-d}\
             reveal that the development of the innermost bubbles\
             is always associated with the disk. Also, without the disk (Fig. \ref{fig__jetI5+ismSeed3-45deg-d}),\
             the outermost bubbles and the turbulent plasma would be oblique,
             indicating that the dense disk is crucial for the production of symmetric Galactic bubbles.}

      \label{fig__jetI5+ismSeed3-45deg}
 \end{figure}

\begin{figure}%
      \begin{subfigure}[t]{0.35\linewidth}%
         \includegraphics[height=8cm]{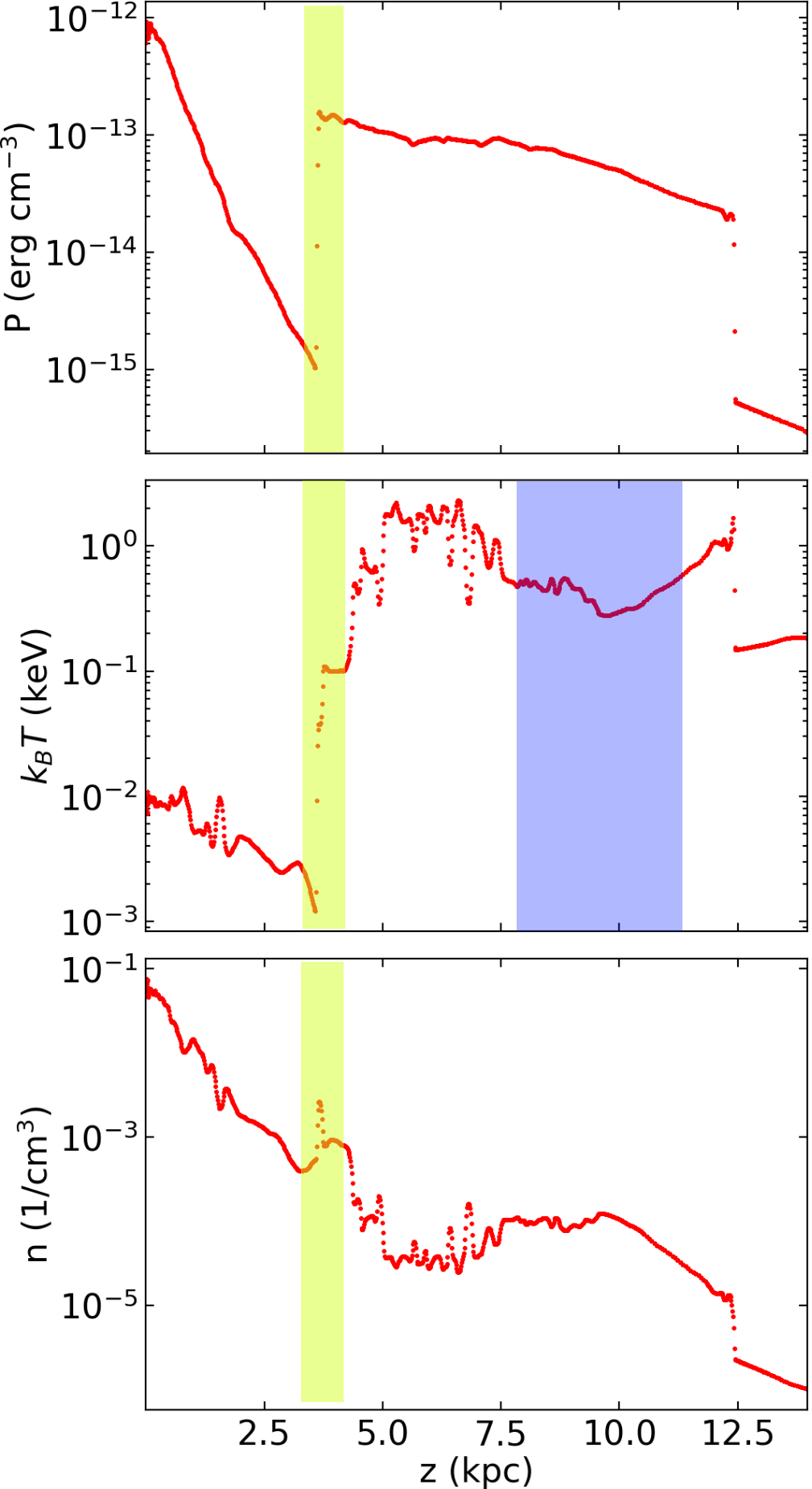}%
         \label{fig__profile-1}%
      \end{subfigure}%
      \hspace{35pt}
      \begin{subfigure}[t]{0.35\linewidth}%
         \includegraphics[height=8cm]{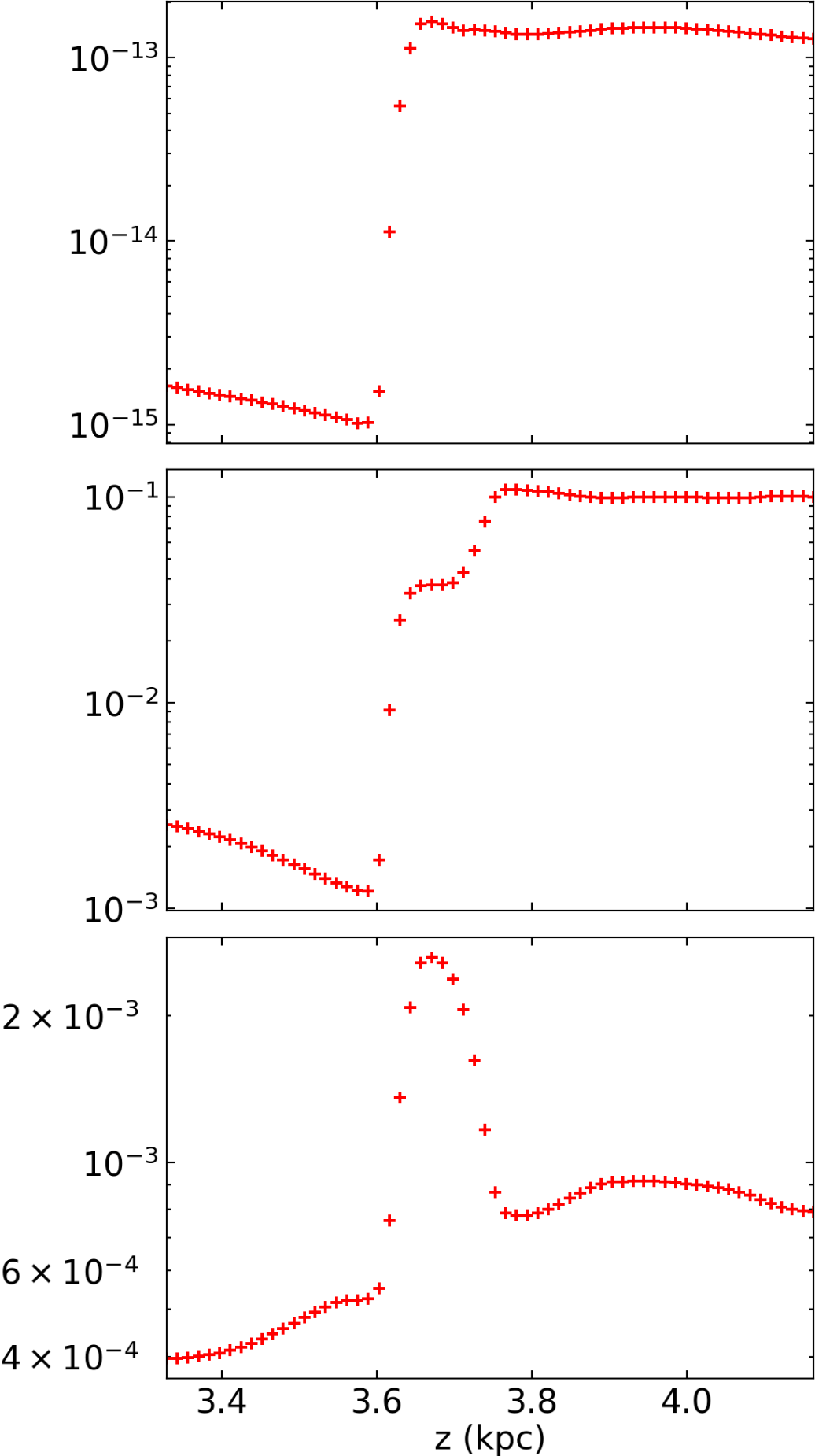}%
         \label{fig__profile-2}%
      \end{subfigure}%
      \caption{
             Left: the profiles of pressure (top), temperature (middle),\
             and number density (bottom)\
             along the positive $z$-axis in Fig. \ref{fig__jetI5+ismSeed3-45deg}.
             Right: the close-up view of the profiles in the yellow band.\
             The sharp pressure jump at $z=3.62$ kpc\
             indicates that the innermost bubbles\
             (dashed box in the top panel in Fig. \ref{fig__jetI5+ismSeed3-45deg-a})\
             are an expanding reverse shock.
      }%
      \label{fig__profile}%
 \end{figure}

 Fig. \ref{fig__jetI5+ismSeed3-45deg-a}-\ref{fig__jetI5+ismSeed3-45deg-d} show\
 the slices of pressure (top), temperature (middle), and number density (bottom)\
 from four different simulations, all ending at $t=12.39$ Myr.\
 The slices pass through the bipolar jet source 
 tilted along $z=-y$ direction. Fig. \ref{fig__jetI5+ismSeed3-45deg-a} demonstrates the fiducial run\
 with the initial condition specified in Section \ref{Methodology}\
 shows that the edge of the outermost bubbles is a forward shock,\
 expanding to 12.5 kpc above and below the Galactic plane,\
 with a semiminor axis of about 6.8 kpc on the plane.\
 The overall extent of the outermost bubbles is comparable to\
 the two spherical objects of a radius of 6-7 kpc estimated by \citet{Predehl2020}\
 for modeling the eROSITA bubbles.\
 The temperature profile (middle panel in the left column of Fig. \ref{fig__profile}) along the positive $z$-axis from the GC indicates that\
 the temperature of the smooth region (purple band in Fig. \ref{fig__profile})\ 
 is around 0.3-0.5 keV, similar to 0.3 keV observed by \citet{Miller2016} and \citet{Kataoka2018}.

\begin{figure}
  \includegraphics[width=\linewidth]{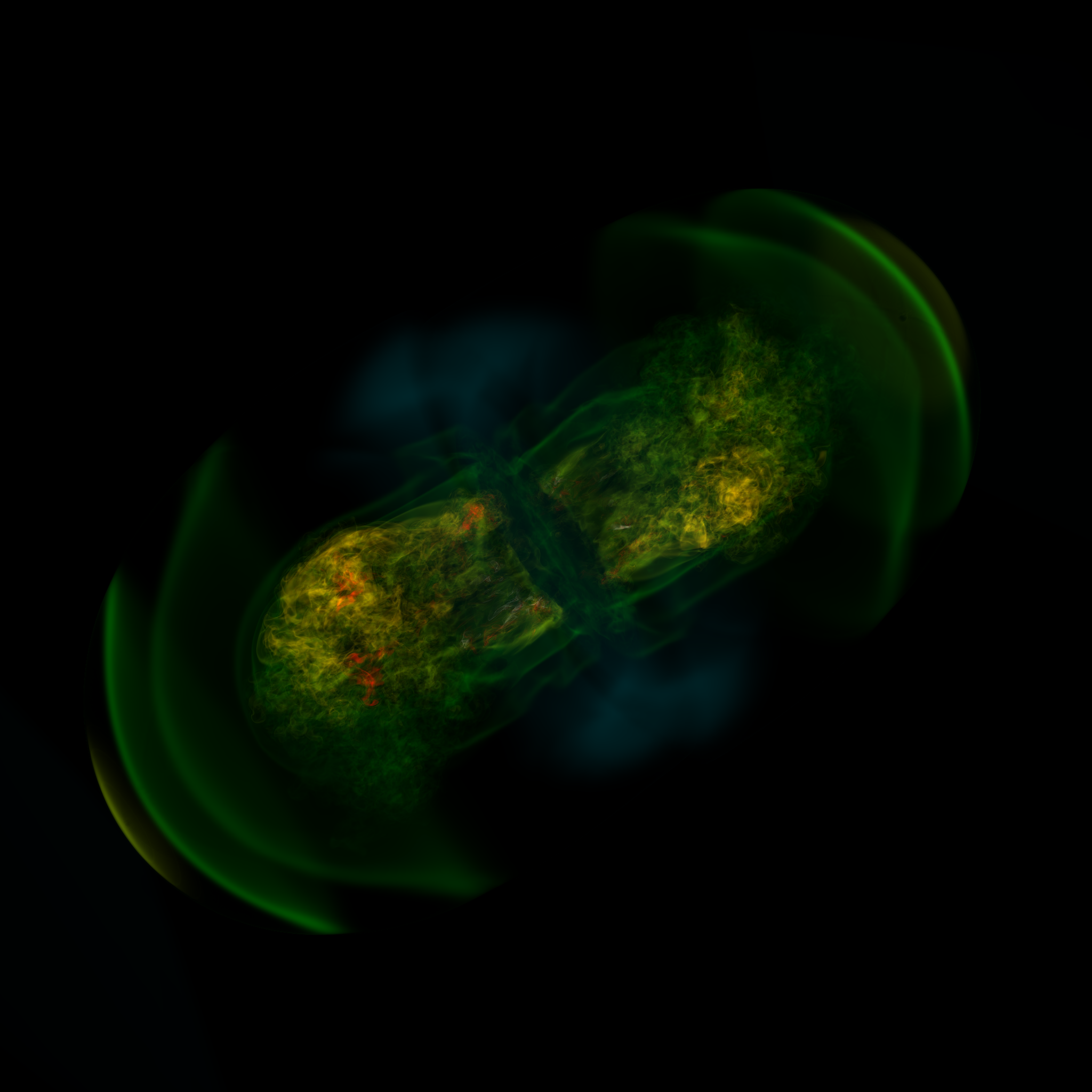}
  \caption
   {
    A 3D isocontour rendering of the temperature of the fiducial run at the present time,
    indicating a turbulent, hot inner structure (\textit{Fermi} bubbles)
    surrounded by a smooth, warm outer shell (eROSITA bubbles).
   }
  \label{fig__rendering.png}
\end{figure}

 Followed by the forward shock is a turbulent and hot plasma extending to a height of $\sim 8$ kpc (Fig. \ref{fig__profile}).\
 The extent of the turbulent plasma approximately agrees with\
 that of the observed \textit{Fermi} bubbles \citep{Su2010}.
 Also, the temperature of the plasma is around 2 keV,\
 comparable to few keV inside the \textit{Fermi} bubbles\
 estimated by observing X-ray absorption lines through the hot\
 gaseous halo along many different sight lines in the sky \citep{Miller_2013}.\
 We also note that the turbulent, hot plasma is in pressure balance with the external medium,\
 suggesting the outer edge of the \textit{Fermi} bubbles\
 is a contact discontinuity rather than a shock \citep{Zhang2020}. A 3D isocontour rendering of the temperature of the fiducial run is shown in Figure \ref{fig__rendering.png}.

\begin{figure}
    \includegraphics[width=\columnwidth]{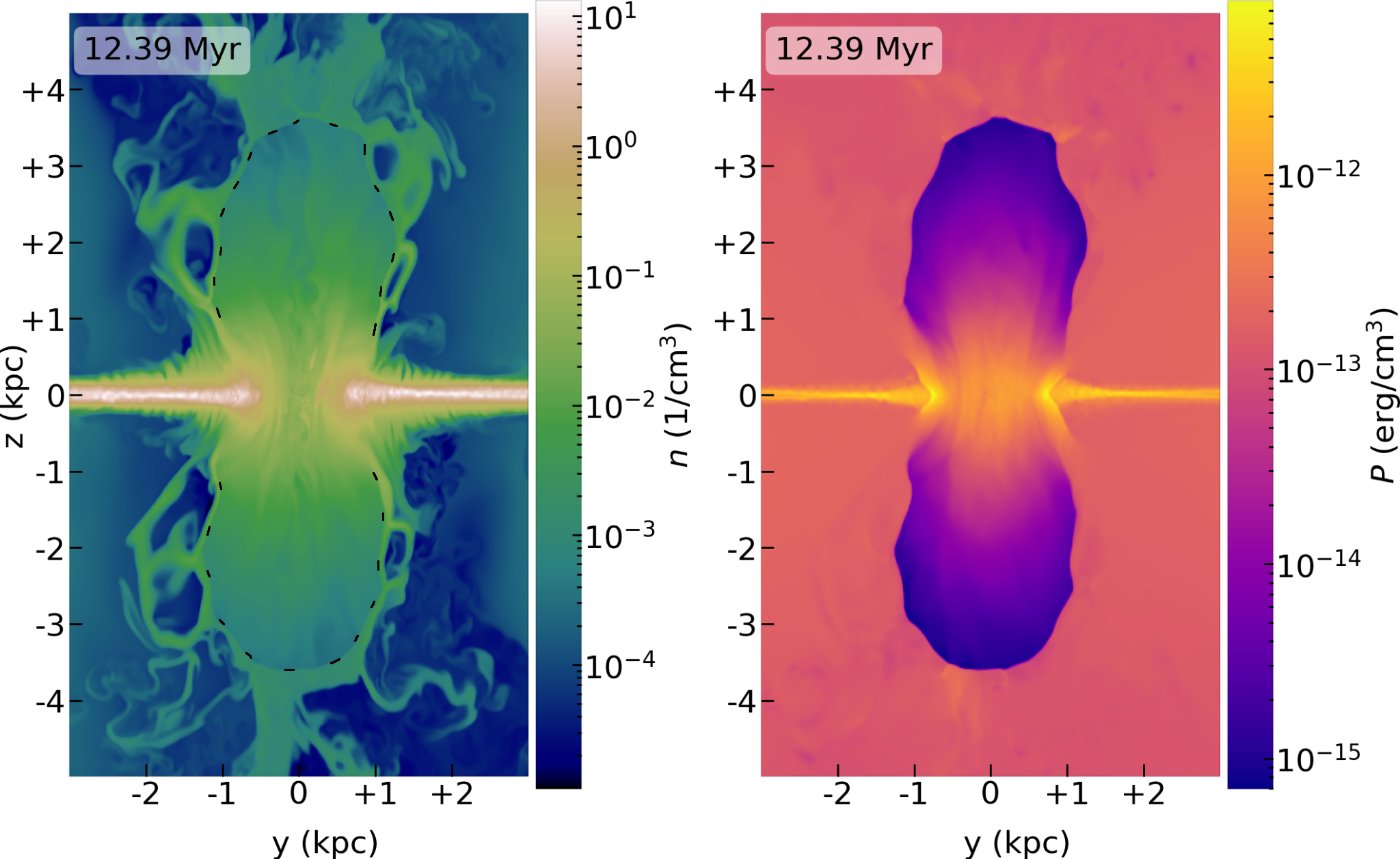}
    \caption{
       Zoom-in images of the density (left) and pressure (right) slices of the innermost bubbles.\
       The high density upstream of the reverse shock requires\
       an even higher density downstream.\
       Continuing outward, the higher density must match with low density further downstream;\
       thus there exists a dense shell\
       (the turbulent region outside the black dashed line in the left panel).
     }
    \label{fig__innerbubbles}
  \end{figure}

 An interesting feature found in our simulations is that there are a pair of innermost bubbles\
 (dashed box in the top panel of Fig. \ref{fig__jetI5+ismSeed3-45deg-a})\
 extending out from the GC on either side of the thin disk.\
 The innermost bubbles are cold (1-10 eV), dense ($10^{-4}$--$10^{-2}$ cm$^{-3}$),\
 and underpressured with respect to the turbulent plasma.
 The close-up view (right column in Fig. \ref{fig__profile}) of the vertical profiles\
 and slices (Fig. \ref{fig__innerbubbles})\
 demonstrate that there is a sharp pressure jump at the edge of the innermost bubbles at $z=3.62$ kpc,\
 indicating that the innermost bubbles are an expanding reverse shock.\
 The high-density upstream of the reverse shock requires\
 an even higher density downstream.\
 Continuing outward, there exists a dense shell before the gas density drops to values further downstream.
 The turbulent plasma is thus confined between the downstream of the reverse shock\
 and the outermost forward shock, resulting in considerable heating of the turbulent plasma.
 These innermost bubbles are probably unrelated to the X-ray chimneys \citep{Ponti2019}\
 and radio bubbles \citep{Heywood2019}\
 due to their enormous difference in length (the total major-axis length of the X-ray chimneys
 and radio bubbles is 320 pc and 430 pc, respectively; however, the innermost bubbles
 is up to 4 kpc in length).

To show what are the key factors for reproducing the symmetry, we performed three additional simulations in different Galactic environments, shown in
Fig. \ref{fig__jetI5+ismSeed3-45deg-b}-\ref{fig__jetI5+ismSeed3-45deg-d}.
Compared with the fiducial run (Fig. \ref{fig__jetI5+ismSeed3-45deg-a}), Fig. \ref{fig__jetI5+ismSeed3-45deg-b} differs by incorporating a smooth disk (the initial density profile is shown in Fig. \ref{fig__initial-density-profile}) in the same atmosphere as the fiducial run, whereas Fig. \ref{fig__jetI5+ismSeed3-45deg-c} differs by using the smooth disk in a uniform atmosphere with a constant temperature $T_{\text{atmp}}=10^6$ K and constant density $\rho_{\text{atmp}}(z=z_{0})$. Fig. \ref{fig__jetI5+ismSeed3-45deg-d} varies by omitting the disk and adopting the stratified atmosphere described by Eq. \ref{isothermal-atmp-density} in the entire domain with a constant temperature $T_{\text{atmp}}=10^6$ K. Note that all cases are initially in hydrostatic equilibrium since Fig. \ref{fig__jetI5+ismSeed3-45deg-b} and \ref{fig__jetI5+ismSeed3-45deg-d} adopt the same gravitational potential specified by Eq. \ref{total-gravitational-potential} as the fiducial run. Fig. \ref{fig__jetI5+ismSeed3-45deg-c} also utilizes this potential, Eq. \ref{total-gravitational-potential}, for $|z|<=z_{0}$, but for $|z|>z_{0}$, a flat potential is used. 

 We compare the clumpy disk (Fig. \ref{fig__jetI5+ismSeed3-45deg-a})\
 with the smooth disk in a stratified atmosphere (Fig. \ref{fig__jetI5+ismSeed3-45deg-b}),\
 showing that the clumpiness of\
 the dense disk has an insignificant effect\
 on the overall dynamics of bubbles. However, the outermost bubbles arising from the smooth disk\
 in a uniform atmosphere (Fig. \ref{fig__jetI5+ismSeed3-45deg-c}) is quasi-spherical,\
 suggesting that the stratification facilitates the elongation of the outermost bubbles significantly. Fig. \ref{fig__jetI5+ismSeed3-45deg-c} and \ref{fig__jetI5+ismSeed3-45deg-d}\
 reveal that the development of the innermost bubbles\
 is always associated with the disk. Also, without the disk (Fig. \ref{fig__jetI5+ismSeed3-45deg-d}),\
 the outermost bubbles and the turbulent plasma would be oblique,\
 indicating that the dense disk is crucial for the production of symmetric Galactic bubbles.\

  \subsection{Morphology and profiles in X-ray}
  \label{X-ray}
  The X-ray emissivity is computed\
  for each computational cell
  using the MEKAL model \citep{Xray-1,Xray-2,Xray-3}\
  implemented in the utility XSPEC \citep{XSPEC}, assuming solar metallicity.\
  The X-ray intensity map is then generated by projecting the emissivities\
  along lines of sight\
  pointing away from the solar position at $(R_{\odot},0,0)=(8,0,0)$ kpc\
  with angular resolutions of 0.5 degrees, where $R_{\odot}$ is the Sun-GC distance. We do not account for any absorption effects, including free-free absorption and photoionization, because adequate modeling would require additional details, such as clumpiness and height of the Galactic disk. Modeling the exact properties of the Galactic disk is beyond the scope of our study and, therefore, is not included in our analysis. We point out that the projections used throughout this paper are \lq perspective\rq,\
  which has the effect of making a distant object appear smaller than the same object at a closer distance,\
  in order to facilitate a reliable interpretation of simulated all-sky map.\
  Also, the observed X-ray emission is contributed by all the gas in the Milky Way halo,\
  which likely extends to a radius of $\sim$250 kpc \citep{halo-radius-1,halo-radius-2},\
  much bigger than our simulation box. Therefore, we first compute the X-ray emissivity\
  from the simulated gas within a radius of 25 kpc away from the GC.
  Then, beyond 25 kpc the gas is assumed to be isothermal with $T=10^6$ K and\
  follows the observed density profile of \citep{temperature-MW} out to a radius of 250 kpc.

\begin{figure*}%
      \begin{subfigure}[t]{\textwidth}%
         \centering%
         \includegraphics[width=0.85\textwidth]{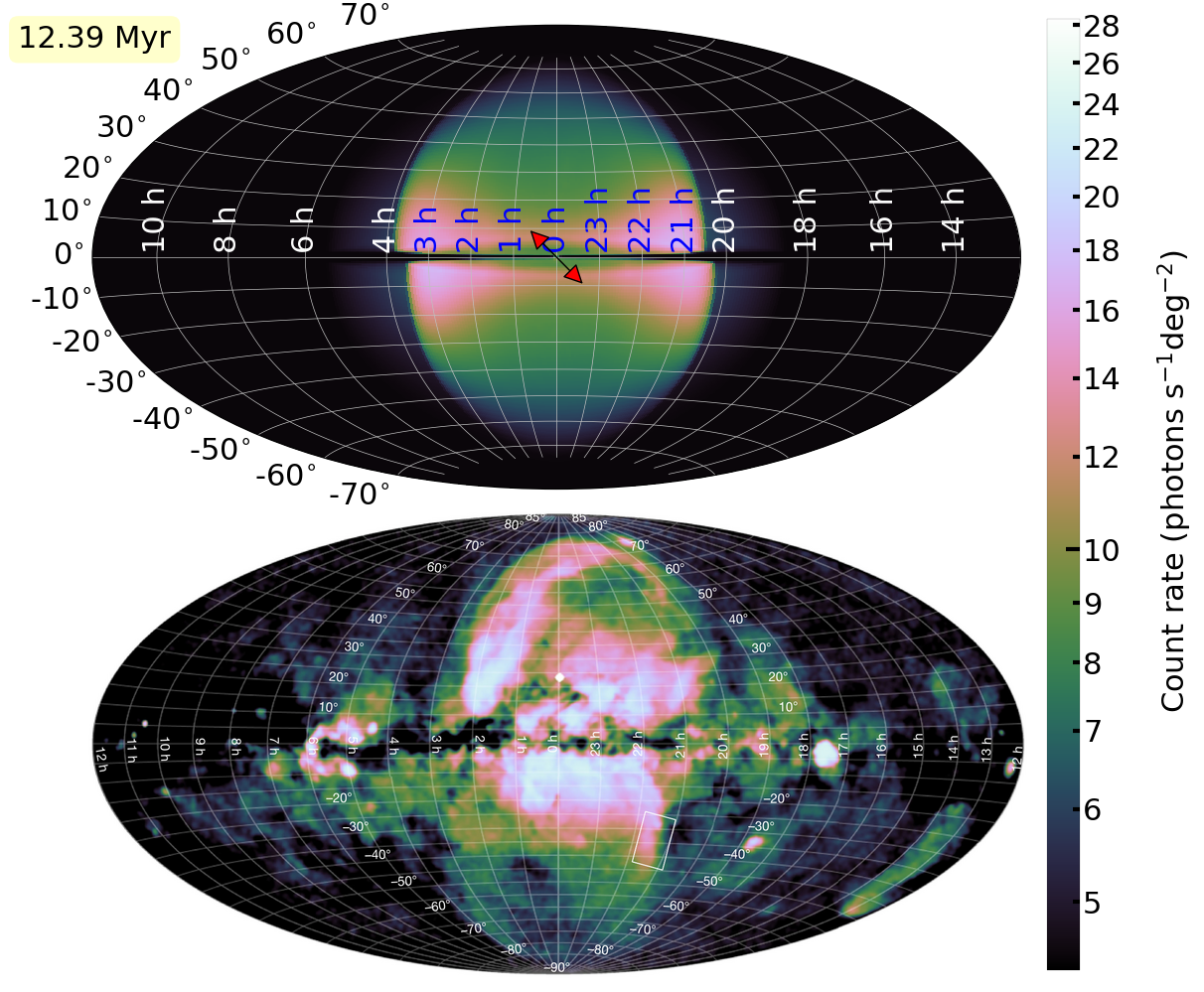}%
         \caption{
               Simulated (top) and observed (bottom; \citealt{Predehl2020}) count rate\
               (photons s$^{-1}$ deg$^{-2}$) in the 0.6--1.0 keV range.\
               Throughout this paper we show sky maps\
               in Galactic coordinates centered on the Galactic center using a Hammer-Aitoff projection.\
               The red arrow at the center of the\
               top panel depicts the direction of the bipolar jets, constantly ejecting at an angle of $45^{\circ}$\
               to the disk normal for the first 0.12 Myr.
         }%
         \label{fig__xray_0.8keV_angle_000}%
         \end{subfigure}%
         \hspace{4pt}%
      \begin{subfigure}[t]{\textwidth}%
         \centering%
         \includegraphics[width=0.85\textwidth]{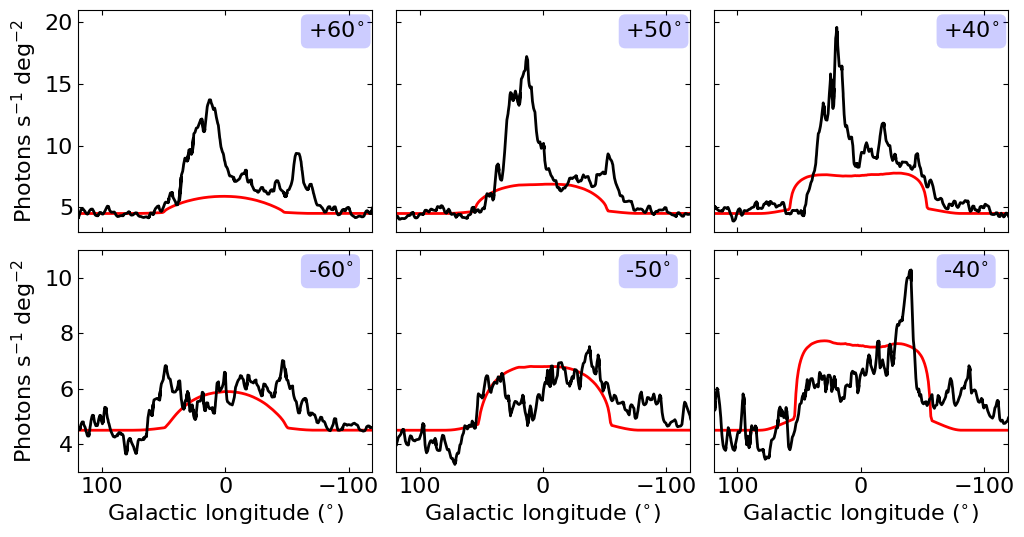}%
         \caption{
               Comparison of the simulated (red) and observed (black; \citealt{Predehl2020}) one-dimensional photon\
               count-rate profiles in the same energy band as in Fig. \ref{fig__xray_0.8keV_angle_000},\
               cut at various Galactic latitudes (as labelled).
         }%
         \label{fig__x-ray-profile-0.8keV-000}%
      \end{subfigure}%
      \caption{}%
      \label{fig_xray-map}
 \end{figure*}

  Fig. \ref{fig__xray_0.8keV_angle_000} shows\
  the comparison between the simulated (top) and observed (bottom) all-sky map\
  in the range 0.6--1.0 keV.\
  In the simulated map, the red arrow at the center represents the direction of the bipolar jets,\
  constantly ejecting at an angle of $45^{\circ}$ to the disk normal between 0--0.12 Myr. Fig. \ref{fig__x-ray-profile-0.8keV-000} displays the simulated X-ray photon count rates as a function of Galactic longitudes (red)\
  in the same energy band as in Fig. \ref{fig__xray_0.8keV_angle_000},\
  cut at various Galactic latitudes (as labelled), compared with the observed profiles (black).


  First, as shown in Fig. \ref{fig__jetI5+ismSeed3-45deg-a},\
  the full width of the outermost bubbles is around 14 kpc,\
  corresponding to a full angular width $2\sin^{-1}(7 \text{ kpc}/R_{\odot})\sim122^{\circ}$,\
  which is as wide as the eROSITA bubbles in the simulated X-ray map\
  (top panel in Fig. \ref{fig__xray_0.8keV_angle_000}).\
  We therefore suggest that the eROSITA bubble shells are a signature of compressed forward shocks\
  that have been driven into the northern and the southern Galactic halo,
  as previously proposed by \citet{Predehl2020} and \citet{Yang2022}. The broad agreement between simulated and observed X-ray maps hints that the full vertical extent of the eROSITA bubbles can be properly formed by an oblique jet within a thin disk of dense ISM.

  Second, we observe that\
  the simulated eROSITA bubbles are not as limb-brightened as the observed ones.\
  A possibility to enhance the X-ray emission is to include shock-accelerated CRs near the shock,\
  in which CRs could increase the compressibility of the fluid,\
  resulting in the enhanced thermal Bremsstrahlung emissivity that is proportional to gaseous density squared. Also, the disagreement of the northeastern bubble is expected as the North Polar Spur, which is a giant ridge of bright X-ray emission that rises roughly perpendicularly out of the plane of the galaxy, might be a superposition of the GC structure and a remnant of the local supernova \citep{berkhuijsen1971galactic,Das2020,Panopoulou2021}. The North Polar Spur is not included in our simulations, whereas analyses based on X-ray data tend to suggest a GC origin \citep{kataoka2018x,sofue2000bipolar,larocca2020analysis}.

  Third, the innermost bubbles\
  shown in Fig. \ref{fig__jetI5+ismSeed3-45deg},\
  even though with high column density, are invisible in the simulated X-ray map as\
  the temperature of the innermost bubbles is around $1$--$10$ eV\
  (see the temperature profile in Fig. \ref{fig__profile}).\
  Consequently, the X-ray emission within the innermost bubbles
  is severely suppressed by the cutoff $\exp\left[-h\nu/k_{B}T\right]$ in the thermal Bremsstrahlung emissivity.\
  This is the reason why the innermost bubbles are unseen in X-rays either through observations or simulations.

\subsection{Gamma-ray and microwave spectra: constraint on the CRe spectral index}
\label{sec:gamma-ray-microwave}
In this section, we obtain the constraint on the CRe spectral index by\
comparing the simulated gamma-ray and microwave spectra\
with the observed spectra of the \textit{Fermi} bubbles \citep{Ackermann2014}\
and the microwave haze \citep{Dobler_2008}, respectively.\

\begin{figure*}
  \includegraphics[width=\linewidth]{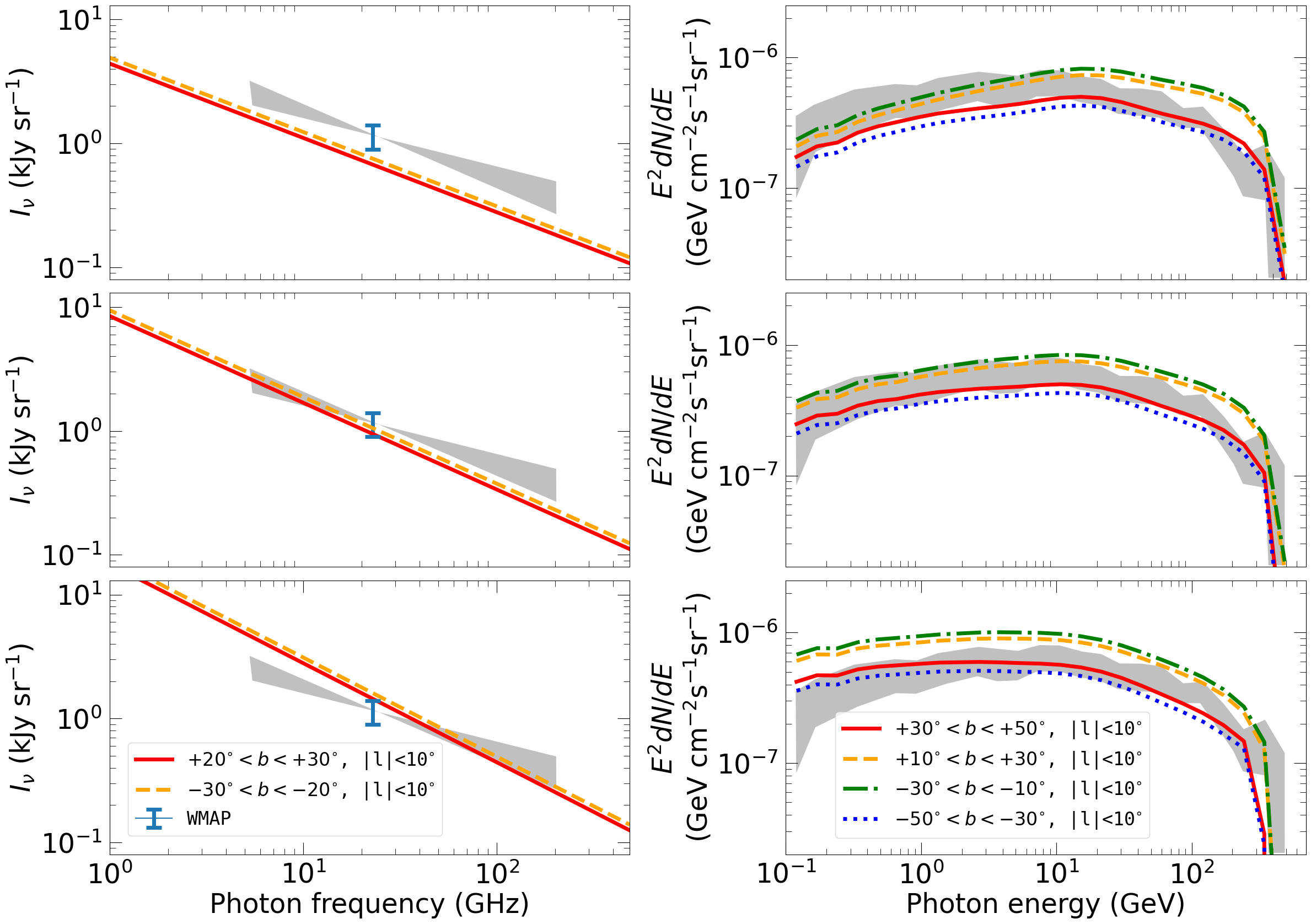}
  \caption{
      Simulated microwave spectra (colored lines in left) averaged over $20^{\circ}<|b|<30^{\circ}$, $|l|<10^{\circ}$.\
      The data point represents the \textit{WMAP} data in the 23 GHz K band and\
      the shaded bow-tie area indicates the range\
      of synchrotron spectral indices allowed for the \textit{WMAP} haze \citep{Dobler_2008}.\
      Simulated gamma-ray spectra (colored lines in right column)\
      of the \textit{Fermi} bubbles calculated for a longitude range of\
      $|l|<10^{\circ}$ for different latitude bins.\
      The gray band represents the observational data of \citet{Ackermann2014}.\
      The row from top to bottom shows the microwave (left) and gamma-ray (right) spectra\
      with CRe spectral index $2.2, 2.4$ and $2.6$, respectively.\
      The CRe cutoff energy is 562.1 GeV in all cases.
  }
  \label{fig__gammaRaySynchtronSpectrum}
\end{figure*}

We assume the leptonic model for the gamma-ray and microwave emission as previous studies have shown that the bubble and haze spectra can be simultaneously produced by the same population of CRe \citep{Su2010, Yang2013, Ackermann2014, Yang2022}. In the leptonic scenario, the gamma-ray and microwave emission come from inverse-Compton (IC) scattering of the interstellar radiation field (ISRF) and synchrotron radiation, respectively.\
Because the evolution of CR spectrum is not modeled in the simulations, we assume that the CRe spectrum follows a power-law distribution:

\begin{equation}
\label{CRe-spectrum}
N(\gamma_{\text{e}})=  \mathbb{C}\gamma_{\text{e}}^{-p_{e}},
\end{equation}

where $\gamma_{\text{e}}$ is the Lorentz factor of the CRe; $p_{e}$ 
is the spectral index of the CRe, assumed to be spatially uniform across the space.
Since $N(\gamma_{\text{e}})$ denotes the number of CRe per unit volume within energy range $\gamma_{\text{e}}$ to $\gamma_{\text{e}}+d\gamma_{\text{e}}$, the normalization constant, denoted as $\mathbb{C}$, can be derived as follows:
\begin{equation}
\mathbb{C}=
\left\{\begin{matrix}
\displaystyle\frac{e_{\text{cr}}}{m_{\text{e}}c^2}\left(\frac{2-p_{e}}{\gamma^{2-p_{e}}_{\text{e,max}}-\gamma^{2-p_{e}}_{\text{e,min}}}\right), \text{ }p_{e} \neq 2,
\\
\displaystyle\frac{e_{\text{cr}}}{m_{\text{e}}c^2}\left[\ln\left(\frac{\gamma_{\text{e,max}}}{\gamma_{\text{e,min}}}\right)\right]^{-1}, \text{ }p_{e}=2,
\end{matrix}\right.
\end{equation}

where $c$ is the speed of light,\
$m_{\text{e}}$ is the electron mass, and $\gamma_{\text{e,max}}$ and $\gamma_{\text{e,min}}$ are the maximum and minimum Lorentz factor of the CRe, respectively.
We stress that the normalization constant varies with space and time since $\mathbb{C}$ is a function of $e_{\text{cr}}$ governed by Eq. (\ref{Ecr evolution}).
We assume that the CRe spectrum, Eq. (\ref{CRe-spectrum}),
ranges from\
$0.5$ MeV ($\gamma_{\text{e,min}}\sim 1$) to $562.1$ GeV ($\gamma_{\text{e,max}}\sim 1.1\times10^{6}$), where $\gamma_{\text{e,max}}$ is motivated by
the observed cutoff gamma-ray energy shown in Fig. \ref{fig__gammaRaySynchtronSpectrum}\
as most of the CRe energy is carried away by the up-scattered photons in the Klein-Nishina limit.

The IC emissivity of the upscattered photons at the energy $\epsilon_{1}$ is computed for\
each computational cell in our simulations\
using the Klein-Nishina IC cross-section \citep{Jones1968,BLUMENTHAL1970}\
to handle the scattering between ultra-relativistic CRe and photons in the ISRF:

\begin{subequations}
  \begin{align}
  &\frac{dE}{dtd\epsilon_{1}dV} =\nonumber\\
               &\frac{3}{4}\sigma_{T}c\mathbb{C}\epsilon_{1}\int^{\epsilon_{\text{max}}}_{\epsilon_{\text{min}}}
               \frac{n(\epsilon)}{\epsilon}d\epsilon\int^{\gamma_{\text{e,max}}}_{\gamma_{\text{e,min}}\left(\epsilon\right)}
               \gamma_{\text{e}}^{-(p_{e}+2)}f(q, \Gamma)d\gamma_{\text{e}},\\
  \nonumber\\
  &f(q, \Gamma) =\nonumber\\
               &2q\ln q+(1+2q)(1-q)+0.5(1-q)\frac{\left(\Gamma q\right)^2}{1+\Gamma q},\\
  &q=\frac{\epsilon_{1}/\gamma_{\text{e}}\
               m_{\text{e}}c^{2}}{\Gamma\left(1-\epsilon_{1}/\gamma_{\text{e}} m_{\text{e}}c^{2}\right)},\\
  &\Gamma=\frac{4\epsilon \gamma_{\text{e}}}{m_{\text{e}}c^2},\\
  &\gamma_{\text{e,min}}(\epsilon)=\
   0.5\left(\frac{\epsilon_{1}}{m_{\text{e}}c^2}+\sqrt{\left(\frac{\epsilon_{1}}{m_{\text{e}}c^2}\right)^2+\
   \frac{\epsilon_{1}}{\epsilon}}\right) \label{gamma-min},
  \end{align}
\label{gammaray-emissivity}
\end{subequations}
where $\sigma_{T}$ is the Thomson cross section;\
$\gamma_{\text{e,min}}(\epsilon)$ is the minimum Lorentz factor of CRe\
that allows the incident photons to be scattered from energy $\epsilon$ to $\epsilon_{1}$;\
$n(\epsilon)$ is the energy distribution of the photon number density in the ISRF given by \citet{Porter2017}, and this distribution varies depending on the Galactic location.\
To obtain the simulated IC emissivities, we perform the double integration in Eq. \ref{gammaray-emissivity} on each cell\
over the range of the CRe Lorentz factor and\
the range of incident photon energy between\
$\epsilon_{\text{min}}=1.13\times10^{-4}$ eV (cosmic microwave background) and\
$\epsilon_{\text{max}}=13.59$ eV (optical starlight).\

The synchrotron emissivity with an isotropic electron pitch angle distribution\
is given by \citet{BLUMENTHAL1970}:

\begin{subequations}
   \begin{align}
      &\frac{dE}{dtd\nu dV} =\nonumber\\
      &\frac{4\pi\mathbb{C}e^{3}B^{0.5(p_{e}+1)}}{m_{\text{e}}c^{2}}\
      \left(\frac{3e}{4\pi m_{\text{e}}c}\right)^{0.5(p_{e}-1)}\
      a(p_{e})\nu^{-0.5(p_{e}-1)},\\
      &a(p_{e})=\nonumber\\
           &\frac{2^{0.5(p_{e}-1)}\sqrt{3}\Gamma\left[\left(3p_{e}-1\right)/12\right]\
                                      \Gamma\left[\left(3p_{e}+9\right)/12\right]\
                                      \Gamma\left[\left(p_{e}+5\right)/4\right]}
      {8\sqrt{\pi}(p_{e}+1)\Gamma\left[\left(p_{e}+7\right)/4\right]},
   \end{align}
   \label{synchrotron-emissivity}
\end{subequations}

where $\Gamma$ is the gamma function, and $B$ is the magnetic field strength defined in\
Eq. \ref{magnetic-field}. For a given longitude and latitude range, the simulated spectra are\
computed by projecting emissivities\
as we project X-ray emissivities in Section \ref{X-ray},\
and then we average the spectra over all the sight lines within a region in the sky.

Fig. \ref{fig__gammaRaySynchtronSpectrum} shows the simulated microwave (left)\
and gamma-ray (right) spectra averaged over the different patches (shown in legends) of the sky.\
The rows from top to bottom show the simulated spectra with different assumptions of the CRe spectral index ($p_{e}$) $2.2, 2.4$, and $2.6$.
To fit the observed spectra,
the CRe energy density ($e_{\text{cr}}$) at the end of the simulation ($t=12.39$ Myr) is scaled by a factor of 0.17, 1.00, and 8.30 for $p_{e}$ of 2.2, 2.4, 2.6, respectively, across the entire simulation domain.
Given that Eq. \ref{Ecr evolution} is linear to the CRe energy density, scaling the CRe energy density is reasonable provided that the resulting CRe pressure remains below the gas pressure throughout the simulation, which we verified to be true.\footnote{After re-scaling, at $t=12.39$ Myr within the \textit{Fermi} bubbles, the ratios of scaled CRe pressure to gas pressure range from $8\times10^{-6}-1.2\times10^{-4}$ for $p_{e}=2.2$, $5\times10^{-5}-10^{-3}$ for $p_{e}=2.4$, and $4\times10^{-4}-7\times10^{-3}$ for $p_{e}=2.6$. At $t=0$, the ratios are also below 1 except for the $p_{e}=2.6$ case; only for this case our simulation, which neglects the dynamical influence of CR pressure, is not fully self-consistent during the early evolution. However, this particular case with $p_{e}=2.6$ is utilized only for comparative purposes.} 

We highlight our findings as follows.\
First, we find that, among the three values of the CR spectral indices assumed, a CRe spectral index of 2.4 (the middle row) provides the best fits to both the observed gamma-ray and microwave spectra. This value is slightly steeper than the best-fit spectral index of $\sim 2.17$ found by \cite{Ackermann2014}. However, we note that our calculations takes into account the 3D variations of the ISRF, whereas the previous constraint was based on the ISRF at a fixed height of 5 kpc from the Galactic plane.
On the other hand, the case with a spectral index of 2.2 provides acceptable fits to the gamma-ray spectrum and the slope of the microwave spectrum, but it falls short to reproduce the normalization of the microwave spectrum. This discrepancy could be resolved if the true magnetic field strength within the \textit{Fermi} bubbles is slightly higher than that in the halo, or possibly considering the contribution from secondary CR electrons generated by a subdominant CR proton component. If that were the case, a spectral index of 2.2 would still be plausible.

\begin{figure}
    \includegraphics[width=0.5\columnwidth]{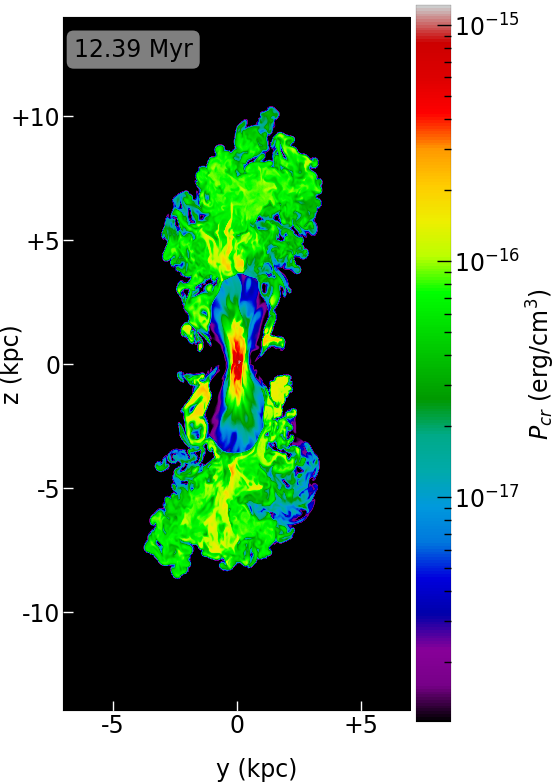}
    \caption{
     The CR pressure slice passing through the jet source at 12.39 Myr.
     Comparison between
     gas pressure (Fig. \ref{fig__jetI5+ismSeed3-45deg}) and cosmic ray pressure
     shows that
     the CR pressure
     is around 5$\times10^{-18}$--$10^{-16}$ erg cm$^{-3}$,
     bringing the CR-to-gas pressure ratio is $5\times10^{-5}-10^{-3}$.
     We therefore stress that ignoring the contribution of CR pressure gradient to the momentum of the gas in Eq. \ref{governing-eq} is reasonable.}

    \label{fig__jetI5+ismSeed3-45deg-CR}
  \end{figure}

Second, the simulated gamma-ray spectra are\
nearly latitude-independent. Note that we have assumed a spatially uniform spectrum for the underlying CRe, and hence the simulated gamma-ray spectra at different latitudes mainly reflect how the 3D distribution of the simulated CR number density (see Fig. \ref{fig__jetI5+ismSeed3-45deg-CR})  is projected into different latitude bins. Overall we find good agreement between the simulated and observed spectra \citep{Ackermann2014}; only the simulated spectrum at high latitudes tends to be slightly dimmer than the lower-latitude spectrum because the optical intensity in the ISRF decays with increasing latitudes.

Third, our assumed range for the CRe spectrum (0.5 MeV to 562.1 GeV) produces gamma-ray spectra with a high-energy cutoff around energies 400--500 GeV,\
consistent with the observed cutoff energy.\
This is expected since\
the upscattered high-energy photons ($\epsilon_{1}\sim450$ GeV) mainly arise from\
the scattering between the relativistic CRe ($\gtrapprox 408$ GeV)\
and optical starlight ($\epsilon \sim 10$ eV).\
Thus, Eq. \ref{gamma-min} can be reduced to $\epsilon_{1}\sim\gamma_{\text{e}} m_{\text{e}}c^2$\
in the Klein--Nishina limit\
$\left(\text{i.e. }\epsilon_{1}\epsilon \gg \left(m_{\text{e}}c^2\right)^2\right)$,\
implying most of the CRe energy is carried away by the upscattered photons.

Finally, the good agreement between the simulated and observed gamma-ray/microwave spectra\
implies that, in the presence of ISRF and magnetic fields, the emission of the \textit{Fermi} bubbles and\
the microwave haze can be produced by the same high-energy electrons\
via IC scattering and synchrotron radiation, respectively. Our results thus provide further support for the leptonic model as previously suggested \citep{Su2010, Ackermann2014, Yang2013, Yang2022}.

\begin{figure*}
  \includegraphics[width=\linewidth]{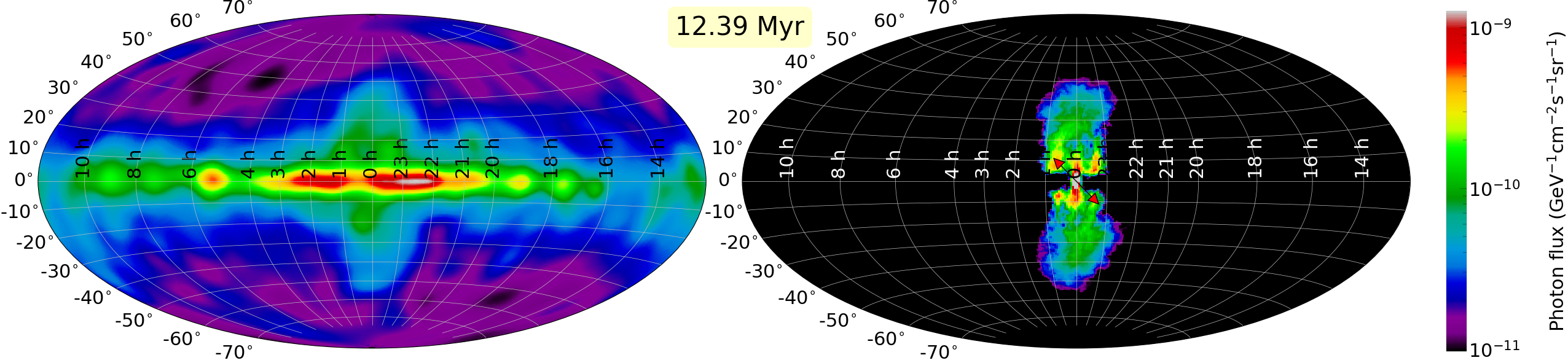}
  \caption{The observed (left; \citealt{Selig2015}) and simulated (right) photon flux\
           in the energy bin $76.8-153.6$ GeV.\
           Note that the left panel is the\
           photon flux of the diffuse component reconstructed by the D$^3$PO\
           algorithm \citep{Selig2015} that analyzes\
           the photon data from the \textit{Fermi} Large Area Telescope \citep{Atwood2009}\
           and removes the contribution from point-like component.
           The red arrow at the center of the right\
           panel depicts the direction of the bipolar jet, constantly ejecting at an angle of $45^{\circ}$\
           to the disk normal in the first 0.12 Myr of the simulation.
  }
  \label{fig__gammaRay-map}
\end{figure*}

 Fig. \ref{fig__gammaRay-map} shows the simulated gamma-ray photon flux with a CRe power-law index 2.4\
 compared with the observed one\
 in the energy bin $76.8-153.6$ GeV.\
 As the eROSITA bubbles,\
 one can see that the symmetric \textit{Fermi} bubbles\
 can also be realized by oblique jets. The extent of the simulated gamma-ray bubbles is also comparable to the observed ones.\
 However, we find that the simulated bubble surface is not as smooth
 as the observed bubbles. The instabilities at the bubble
 surface may be suppressed by the magnetic draping
 effect \citep{Lyutikov2006,Yang2012} if magnetic fields were
 included in the simulations. With magnetic draping, the sharp edges of the observed bubbles \citep{Su2010, Ackermann2014} could also be explained by anisotropic CR diffusion along field lines \citep{Yang2013}.

 As can be seen in Fig. \ref{fig__gammaRay-map}, though the overall size of the simulated gamma-ray bubbles is comparable to that of the observed ones, the gamma-ray intensity does not appear to be as uniform as originally found in \citet{Su2012}.
 As discussed above, the gamma-ray intensity is slightly
 higher close to the Galactic plane due to the stronger
 radiation field at lower latitudes. However, this level
 of brightness variations appears to be consistent with the later
 observational data of \citet{Ackermann2014} and \citet{Selig2015}, which shows that there are some substructures in the gamma-ray intensity distribution within the bubbles.

 For completeness, we show the simulated CR energy density at 12.39 Myr in Fig. \ref{fig__jetI5+ismSeed3-45deg-CR}. The comparison between\
 Fig. \ref{fig__jetI5+ismSeed3-45deg}\
 and\
 Fig. \ref{fig__jetI5+ismSeed3-45deg-CR}\
 shows that the CR pressure\
 is around $5\times10^{-16}$--$10^{-14}$ erg cm$^{-3}$,\
 corresponding to a CR-to-gas pressure ratio of 0.005--0.1,
 similar to 0.15 at the beginning of the simulation.
 We therefore stress that\
 ignoring the contribution of CR pressure gradient to the momentum of the gas\
 in Eq. \ref{governing-eq} is reasonable.

\section{Discussion}
\label{sec:discussion}
The success of the model hinges on the major assumption that
the tilted jets dissipate almost all their kinetic energy within the cold ISM disk without directly penetrating through it.
To find an optimal parameter set satisfying this assumption, we explore jet parameters, such as density, velocity, and temperature
at varying jet tilt angles of $0^{\circ}, 45^{\circ},$ and $90^{\circ}$, as detailed in the Appendix \ref{appendix_A}.
The findings indicate that 1) higher jet density and velocity cause the jets to penetrate the disk and form asymmetric bubbles; 2) lower jet density and velocity result in cooler post-shock regions, inconsistent with observations; 3) alternating the jet temperature --- increasing it leads to hotter post-shock regions, while decreasing it results in cooler ones --- also yields results that are inconsistent with observations; 4) both lying and vertical jet orientations produce dimmer $\gamma$-ray bubbles at high latitudes than the observed ones.
The ideal jet parameters emerged as a temperature of $k_{B}T=1.72$ MeV, a density of $\rho=10^{-26}$ g/cm$^3$, and a four-velocity of $\beta\gamma=0.6$ at $45^{\circ}$ inclination.
Note that these optimal parameters are derived by assuming a clumpy cold disk with a constant half-height of $z_0 = 100$ pc; however, the actual disk height may vary with the distance to the Galactic center.
Closer to the GC, the cold disk is even thinner -- the half-height of the central molecular zone is about 30 pc, and the central circumnuclear gaseous disk may be even thinner.
Under these conditions, it may be easier for the jets to penetrate through the cold disk,
producing tilted \textit{Fermi} and eROSITA bubbles inconsistent with observations; a lower jet density or velocity may be needed to remedy such cases.

Our modeled gamma-ray and microwave spectra assumed that
the underlying CRe spectrum is spatially uniform with a  spectral index of 2.4;
however, the high energy CRe should severely suffer from synchrotron and IC losses,
during their passage through the magnetic and radiation fields, respectively, within the Galaxy. The typical synchrotron and IC cooling time scale of high energy ($\sim100$ GeV) CRe in the Milky Way is $\sim$ 1 Myr \citep{Yang2017}, ten times shorter than the formation time of 12 Myr as suggested by our simulations. Therefore, the CRe generating the gamma-ray emission would need to be re-accelerated
by in-situ acceleration mechanisms such as shocks or turbulence.
Since the forward shock is rather far away from the gamma-ray bubbles,
re-acceleration is more likely associated with turbulence \citep{Mertsch2011,Mertsch2019},
possibly in balance with the IC and synchrotron cooling. We will investigate the
competition between stochastic acceleration and radiative cooling in future work. Overall, our results imply that in the oblique-jet scenario,
regardless of what the true re-acceleration and cooling mechanisms are,
the CRe spectrum at the present time has to be spatially uniform with
a spectral index of 2.4 in order to fit the observed spectra.

Recent simulations by \cite{Sarkar_2023} and \cite{Dutta24} have also investigated the oblique-jet scenario in terms of generating the Galactic bubbles. Our results agree with their findings that, in the oblique-jet scenario, the jets must have significant dissipation within the Galactic disk in order to produce symmetric Galactic bubbles as the observed ones. However, they found that only jets with super-Eddington power ($L_{\rm j} \gtrsim 5\times 10^{44}$ erg s$^{-1}$) could dissipate sufficient energy and produce symmetric bubbles, while lower-power jets tend to penetrate the disk and produce asymmetric bubbles. In contrast, our simulations show that relatively low-power jets ($L_{\rm j} = 3.2\times 10^{42}$ erg s$^{-1}$) could still dissipate a significant amount of energy before leaving the disk and generate symmetric bubbles as observed (see also \cite{Mukherjee2018} or M18). In the Appendix of \cite{Sarkar_2023}, they attributed their different results from M18 to numerical resolution. However, our resolution is $\sim$10 times greater and we still find significant jet dissipation as in M18. Therefore, we suspect that the difference in the results may be due to that \cite{Sarkar_2023} performed non-relativistic hydrodynamic simulations with an initial jet velocity of $0.1$c, whereas our simulations (as well as M18) have used SRHD simulations (our initial jet velocity is set by $\beta \gamma = 0.6$, corresponding to $v\sim 0.5$c). Thus for the same jet luminosity, their jets are relatively denser ($\rho_{\rm jet}/\rho_{\rm amb}\sim 0.1$) and slower, which have higher momentum and tend to penetrate the disk. On the other hand, our relativistic jets are lighter ($\rho_{\rm jet}/\rho_{\rm amb}=10^{-3}$) and faster, and hence are easier to dissipate due to the interaction with the dense disk. Of course, the true initial jet velocity and density contrast on the scale of injections are uncertain. However, our simulations show that the oblique-jet model could be plausible at least for some reasonable range of jet parameters, and that the range of jet luminosity allowed is dependent upon the details of how the jet injections are modeled.

Our simulations demonstrate that it is plausible to relieve the assumption of jet inclinations in terms of forming symmetric Galactic bubbles. However, we stress that there are several differences between this work and previous leptonic jet models that have assumed vertical jets \citep{Guo2012, Yang2012, Yang2013, Yang2017, Yang2022}. In the previous leptonic jet models, the required jet power is near-Eddington, and thus the dissipation within the disk should be minor. Subsequently, the bubbles are formed from the lateral expansion when the ram pressure of the jets balances the external pressure, similar to the formation of cocoons of radio galaxies \citep[e.g.,][]{Begelman1989, Bromberg2011}. Because it takes only $\sim 2-3$ Myr for the Galactic bubbles to form, the advantage is that the jets could transport the CRe before they cool and reproduce the uniformly hard gamma-ray spectrum \citep{Yang2022, Yang2023}. The time scales and energetics are consistent with those inferred from elevated ionization signatures in the Magellanic Stream \citep{BlandHawthorn2019}. The drawback, though, is that it requires the jets to be perpendicular to the Galactic plane, which is either coincidental or requires efficient alignment between the black hole spin and the Galactic disk \citep[e.g.,][]{Fiacconi2018}. On the other hand, if the jets were initially tilted with respect to the rotational axis of the Galaxy, then significant jet dissipation would be required to turn the kinetic energy of the jets into thermal energy. In this scenario, the bubbles are then inflated due to their internal pressure rather than the ram pressure. The advantages of the oblique-jet scenario are that it relaxes the assumption about the jet direction, and that only a moderate jet power is required (though the jet power is not uniquely determined, as influenced by the exact Galactic halo profiles used as well as numerical resolution as discussed in the last paragraph). Since it takes $\sim 12$ Myr for the bubbles to form, which is greater than the CRe cooling times, it remains to be seen whether the spatially uniform hard CR spectrum could be explained by taking into account turbulence re-acceleration within the bubbles.

\cite{Zhang2020} proposed that the \textit{Fermi} bubbles are the product of a forward shock driven by a past jet source at the GC vertical to the Galactic plane. This model supports the idea that the \textit{Fermi} bubbles are directly linked to the dynamical impact of the jet event that imparts significant energy into the surrounding medium. The forward shock, in this scenario, sweeps up and heats the interstellar gas, creating the X-shaped structure near the GC as observed by ROSAT \citep{BlandHawthorn03}. 
 Following this line of reasoning, the eROSITA bubbles ought to be interpreted as a phenomenon unrelated to the \textit{Fermi} bubbles, possibly attributed to a prior energetic outburst. This perspective is different from this work and previous leptonic jet models \citep{Yang2022} that seek to explain both the \textit{Fermi} and eROSITA bubbles as driven by a single event. 

\section{Conclusions}
\label{Conclusions}
In this work, we introduce a thin, dense disk composed of clumpy ISM\
to stall and thermalize the oblique jets for an outburst event from
the central SMBH in the Milky Way Galaxy 12 Myrs ago.
We investigate the properties of the Galactic bubbles and the microwave\
haze using 3D SRHD simulations of CR jet injections from\
the SMBH assuming the leptonic model. The important findings are summarized as follows.


\begin{itemize}

\item The development of the expanding forward-reverse shock pair is always associated with the dense disk.
In the absence of the disk, the reverse shock is absent,
indicating the inclusion of the disk is critical for forming the innermost bubbles.

\item The forward-reverse shock pair heats the turbulent plasma considerably ($\sim2$ keV).
There exists a
dense shell immediately downstream of the reverse shock, a situation reminiscent of a supernova shell.

\item eROSITA bubbles coincide with the forward shock front
      originally driven by short-lived bipolar jets
      for a duration of 0.12 Myr, where the bubbles later significantly
      expanded into the stratified atmosphere to reach the present height of 12 kpc.
      The overall extent of the simulated X-ray bubbles is comparable to
      that of the eROSITA bubbles, though not as limb-brightened.
      Future models including shock-accelerated CRs may help to resolve this issue by
      increasing the compressibility of the fluid and
      enhancing the thermal Bremsstrahlung emissivity
      at the edge of the X-ray bubbles.
\item Downstream of the reverse shock is filled with hot ($\sim2$ keV) and highly turbulent
      plasma brought from the disk. The interface between the downstream materials of reverse and
      forward shocks lies a contact discontinuity, which corresponds to the edge of the \textit{Fermi} bubbles.
      The surface of the simulated bubbles is
      not as smooth as the observed ones; the inclusion of magnetic fields in the future may help suppress
      the instabilities at the bubble surface due to the magnetic draping effect.
\item Assuming a power-law CRe energy spectrum ranging from 0.5 MeV to 560 GeV,
      where the spectrum is spatially uniform,
      we showed that the observed gamma-ray and microwave spectra
      can be simultaneously reproduced. The best-fit CRe power-law index is found to be 2.4 given our model assumptions. A spectral index of 2.2 could also be plausible if the amount of the microwave emission could be boosted by, e.g., a larger magnetic field than what was adopted, or additional contribution from secondary CR electrons.
\item The elapsed time of 12 Myr is 10 times longer than
      the typical synchrotron and IC cooling time scale of
      high energy ($\sim100$ GeV) CRe. Thus, re-acceleration of CRe by shocks or turbulence must be considered
      in this model. Since the forward shock is rather far away from the gamma-ray bubbles,
      stochastic acceleration of CRs by turbulence appears to be more plausible.\
      We will investigate the competition between stochastic acceleration and radiative cooling
      in future work.
\item The Galactic bubbles are observed nearly symmetric about the Galactic plane albeit\
      the bipolar jets are oblique with respect to the disk normal.\
      We showed that the inclusion of the dense ISM disk (regardless of its clumpiness)
      is an essential ingredient for producing the symmetric Galactic bubbles when the jets are oblique. The influence of varied jet parameters is explored in detail in the Appendix \ref{appendix_A}. An optimal combination of jet parameters is inferred; however, we caution that they could be influenced by the assumptions of our initial setup (e.g., constant height of the Galactic disk, lack of radial dependence of the Galactic halo profile, etc). Nevertheless, \
      the broad agreement between the simulated and observed multi-wavelength features\
      demonstrates that oblique failed jets are a plausible
      scenario for the formation of the Galactic bubbles,
      which relieves the caveat of earlier jet models where jets need to be vertical.

\end{itemize}


\section{Acknowledgements}
The authors thank Mateusz Ruszkowski and Ellen G. Zweibel for insightful comments.\
We thank Henry Zovaro for discussions about the innermost bubbles\
during the final stages of this work.\
We thank Peter Predehl for providing the range of observed X-ray intensity in Fig. \ref{fig__xray_0.8keV_angle_000}.\
We further thank Chen Chun-Yen for providing the code module solving the cosmic ray convection equation.
A part of the simulations are performed and analyzed using computing resources\
operated by the National Center for High-Performance Computing (NCHC).
HYKY acknowledges support from\
Yushan Scholar Program of the Ministry of Education of Taiwan\
and\
Ministry of Science and Technology of Taiwan (MOST 109-2112-M-007-037-MY3; NSTC 112-2628-M-007-003-MY3).
HS acknowledges funding support from the Yushan Scholar Program No. NTU-111V1201-5 and the NTU Academic Research-Career Development Project under Grant No. NTU-CDP-111L7779,\
sponsored by the Ministry of Education, Taiwan.\
This research is partially supported by the MOST under grants 107-2119-M-002-036-MY3\
and NSTC 111-2628-M-002-005-MY4, and the NTU\
Core Consortium project under grants NTU-CC-108L893401 and\
NTU-CC-108L893402. TC acknowledges support from National Science and Technology Council (NSTC) under grants 111-2112-M-002-031.

\appendix
\section{jet parameter variations}
\label{appendix_A}

\begin{figure}
  \includegraphics[width=\textwidth]{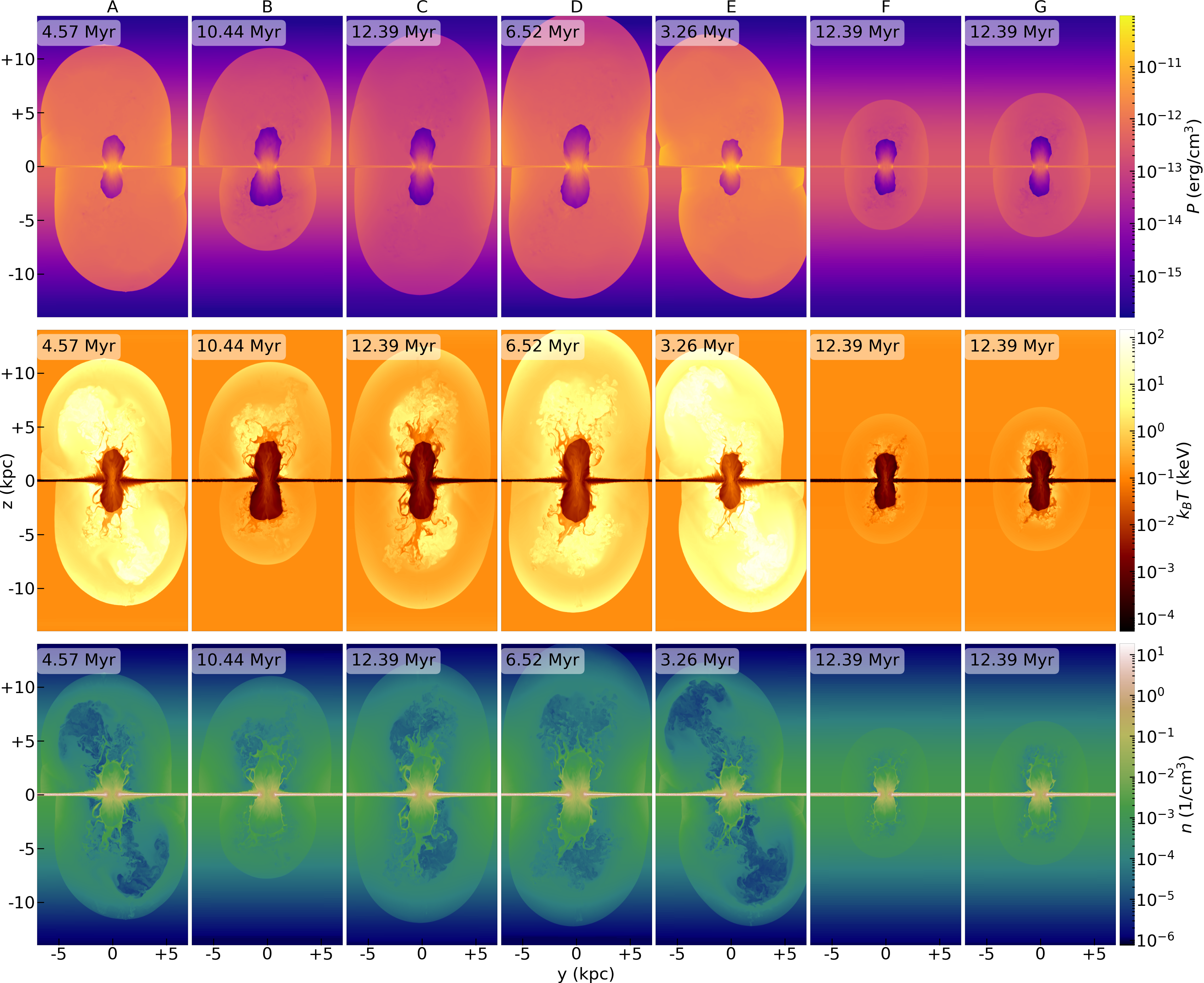}
  \caption
   {
    Simulated gas distributions based on the parameters of the jet in Table. \ref{table-jet-parameters} for
    jet sources with the same inclination angle of 45$^{\circ}$.
   }
  \label{fig__slice-compare}
\end{figure}

Fig. \ref{fig__slice-compare} shows the
simulated Galactic bubbles for a jet source with the same inclination angle of 45 degrees,
but with different jet parameters, including gas temperature, density, and initial velocity. The parameter range explored in this study,
including the fiducial run (C),\
is listed in Table \ref{table-jet-parameters}. The simulated gas pressure, temperature, and density distributions for these simulations are shown in Fig. \ref{fig__slice-compare}. As discussed in the second paragraph of \S~\ref{sec:discussion}, since the simulation results (i.e., whether jets could dissipate within the disk or not) are quite sensitive to the details of jet injections (e.g., size of the jet source, the average density of the disk, simulation resolution, etc), the range of jet parameters that could successfully produce symmetric bubbles may vary when the simulation details are different, and the readers should be cautious when quoting these parameters at face values. Also, because it is computationally expensive to cover all the possible parameters, the parameter space explored here is by no means complete. Instead, our main goal here is to show the qualitative trends when some of the parameters are adjusted.

We observe that higher density (F$\rightarrow$C$\rightarrow$E)\
or velocity (G$\rightarrow$C$\rightarrow$A) of the source\
can enhance the intensity of the shocks driven into the atmosphere.\
As a result, the post-shock region becomes hotter,\
and the bubbles show greater asymmetry around the GC.
This results from the fact that the denser/faster jets are more likely to\
penetrate the thin disk in a short time,\
which prevents the jets from having enough time to interact with the disk and alter their orientations\
before being stretched vertically by the stratified atmosphere.

Also, we found that increasing the source temperature (B$\rightarrow$C$\rightarrow$D)\
not only improves the symmetry of the bubbles but also increases the temperature
inside the bubbles at the present time. This result implies that a high-temperature source
helps the jets ablate the clumpy dense gas away and pour into the atmosphere
through the intercloud channels of the clumpy disk. In contrast,
if the source temperature is not high enough,
the jets cannot effectively ablate the high-density clouds within 12 Myr,
resulting in the formation of asymmetric bubbles (case B).

\begin{figure*}
  \includegraphics[width=0.5\textwidth]{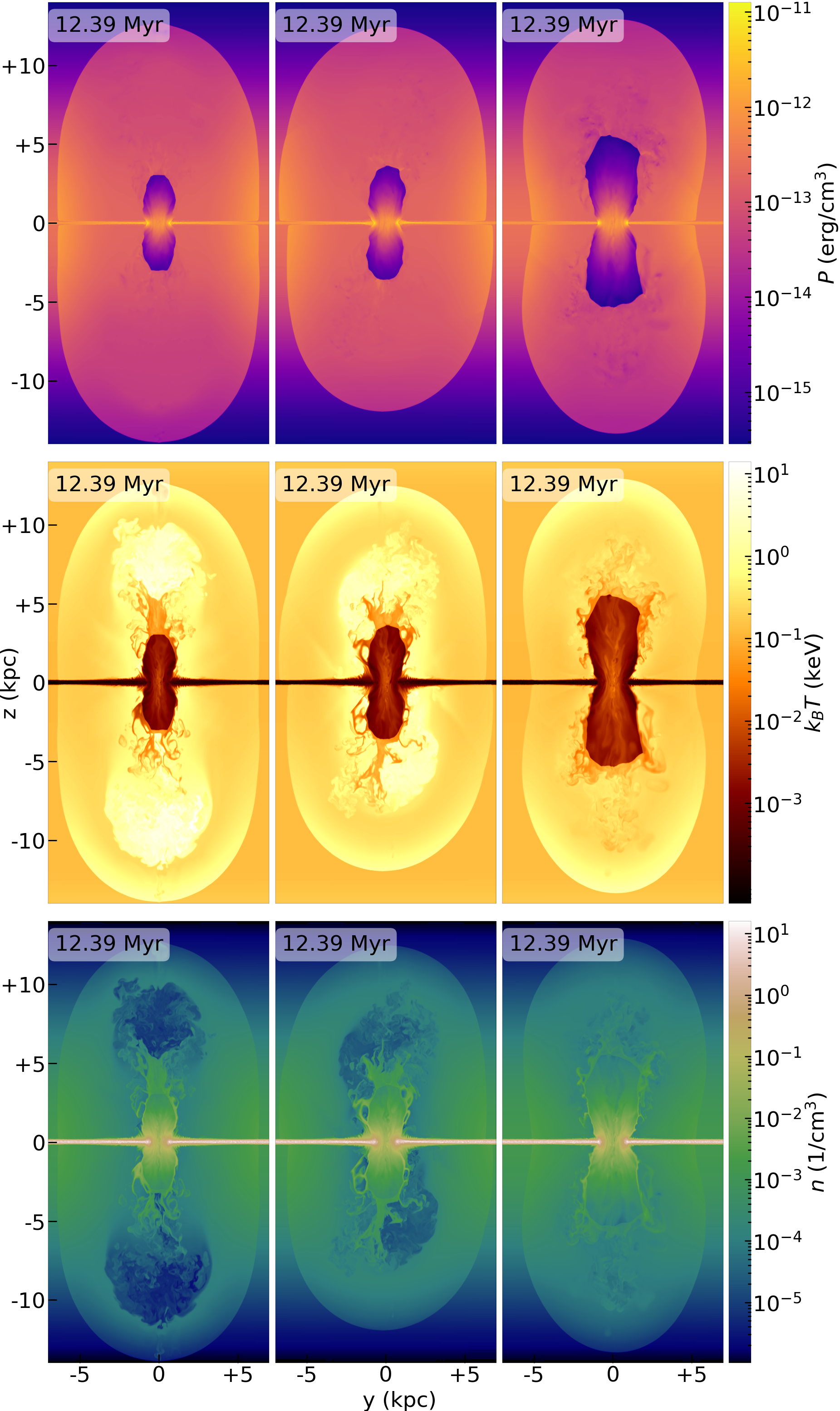}
  \caption
   {
    Simulated gas distributions for the case C in Table. \ref{table-jet-parameters}
    with inclination angles
    0$^{\circ}$ (left), 45$^{\circ}$ (middle), 90$^{\circ}$ (right) with respect to the $z$-axis.
   }
  \label{fig__slice-compare-angle}
\end{figure*}

\begin{figure*}
\includegraphics[width=\textwidth]{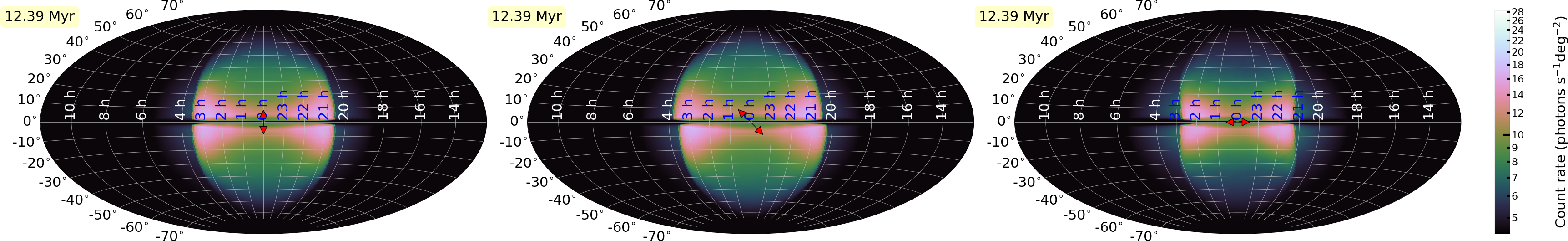}
  \caption
   {
   Simulated X-ray bubbles for the case C in Table. \ref{table-jet-parameters}
    with inclination angles
    0$^{\circ}$ (left), 45$^{\circ}$ (middle), 90$^{\circ}$ (right) with respect to the $z$-axis.
   }
\label{fig__fig__xray-00-45-90-degrees}
\end{figure*}

\begin{figure*}
\includegraphics[width=\textwidth]{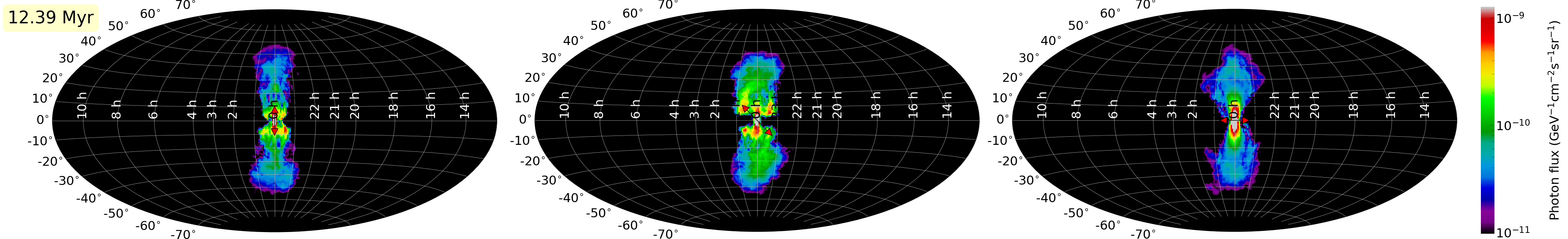}
  \caption
   {
      Simulated $\gamma$-ray bubbles for the case C in Table. \ref{table-jet-parameters}
    with inclination angles
    0$^{\circ}$ (left), 45$^{\circ}$ (middle), 90$^{\circ}$ (right) with respect to the $z$-axis.
   }
\label{fig__gamma-ray-00-45-90-degrees}
\end{figure*}

\begin{figure*}
\includegraphics[width=0.7\textwidth]{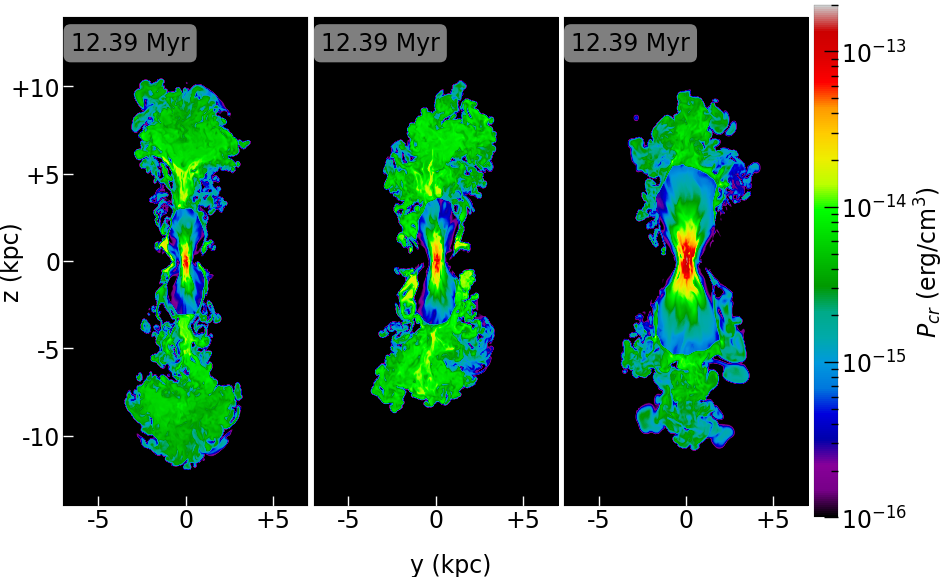}
  \caption
   {
      Simulated CR distributions for the case C in Table. \ref{table-jet-parameters}
    with inclination angles
    0$^{\circ}$ (left), 45$^{\circ}$ (middle), 90$^{\circ}$ (right) with respect to the $z$-axis.
   }
\label{fig__jetI5+ismSeed3-00-45-90deg-CR}
\end{figure*}

Fig. \ref{fig__slice-compare-angle}, \ref{fig__fig__xray-00-45-90-degrees}, \ref{fig__gamma-ray-00-45-90-degrees}, and \ref{fig__jetI5+ismSeed3-00-45-90deg-CR} present the simulated distributions of gas, X-ray, $\gamma$-ray, and CRs for case C. Panels from left to right correspond to different jet inclination angles (0$^{\circ}$, 45$^{\circ}$, and 90$^{\circ}$) relative to the $z$-axis. The overall morphology of the Galactic bubbles are similar among the three cases, indicating that the jet inclination angle has a negligible impact on the orientation of bubbles for the parameter space investigated in this study.

Despite the similarities of the bubble morphology, there are subtle differences of the observable features among the three cases. The bottom row of Fig. \ref{fig__slice-compare-angle} shows that the larger the inclination angle, the denser the gas within the region between the outermost and innermost bubbles.
In addition, the outermost bubbles produced by lying jets (rightmost column in Fig. \ref{fig__slice-compare-angle}) are the narrowest compared to others, resulting in shorter path lengths within the X-ray bright regions inside the outermost bubbles. Consequently, the X-ray bubbles arising from lying jets (the rightmost panel in Fig. \ref{fig__fig__xray-00-45-90-degrees}) appear dimmer than the other two cases. However, as the extent of X-ray emission is larger than the Galactic bulge, please note that the simulated X-ray map may be different from the results obtained by the self-consistent Galactic model, involving radial variation of gravitational potential and the centrifugal force introduced by the rotation of the Galaxy.

The $\gamma$-ray maps (Fig. \ref{fig__gamma-ray-00-45-90-degrees}) reveal that the case produced by the $45^{\circ}$ jets (middle panel) produces a $\gamma$-ray intensity distribution that is most consistent with the nearly flat distribution of the observed {\it Fermi} bubbles compared to other inclinations. Specifically, the case with $45^{\circ}$ jets produces brighter $\gamma$-ray emission than the other two inclinations. This could be attributed to the two facts: (1) the lying jets provide less vertical gas velocity to carry the CRs away from the GC. Therefore, a significant portion of the injected CRs remains proximate to the GC at the present time, giving rise to dimmer $\gamma$-ray bubbles at high latitudes (rightmost panel in Fig. \ref{fig__gamma-ray-00-45-90-degrees}); (2) the vertical jets offer rapid vertical gas flows before quenching, causing CRs to adiabatically expand (Fig. \ref{fig__jetI5+ismSeed3-00-45-90deg-CR}). As a result, the CR energy density decreases at high latitudes, leading to dimming of the $\gamma$-ray bubbles as well (leftmost panel in Fig. \ref{fig__gamma-ray-00-45-90-degrees}).

In summary, we conclude that the initial jet inclination angle has a minor influence on the ability to generate symmetric bubbles for the same set of jet parameters. However, tilted jets (e.g., $45^{\circ}$) could mitigate several issues: (1) preventing the outermost bubbles from being too narrow, which would dim the X-ray bubbles, (2) avoiding a significant pile-up of CRs near the GC, and (3) averting the dilution of CRs at high latitudes due to fast adiabatic expansion. The last two issues would result in non-uniform $\gamma$-ray intensity distributions, at odds with the observed {\it Fermi} bubbles.

\begin{table*}
\raggedright
\caption{The list of runs showing different injected jet parameters.\
         The fiducial run (C) has been pointed out by an asterisk in the list. }
\label{table-jet-parameters}
\begin{tabular}{@{}llllllll@{}}
\toprule[1pt]\midrule[0.3pt]
                  & A          & B          & C*         & D           & E          & F          & G          \\ \midrule
$k_{B}T$ (MeV)    & 1.720      & 0.034      & 1.720      & 86.100      & 1.720      & 1.720      & 1.720      \\
$\rho$ (g/cm$^3$) & $10^{-26}$ & $10^{-26}$ & $10^{-26}$ & $10^{-26}$  & $10^{-25}$ & $10^{-27}$ & $10^{-26}$ \\
$\beta\gamma$     & 1.20       & 0.60       & 0.60       & 0.60        & 0.60       & 0.60       & 0.30       \\ \midrule
\end{tabular}
\end{table*}

\section{Cosmic-Ray Implementation and Tests}
\label{appendix_B}
This appendix provides an overview of the full implementation of cosmic rays (CRs) in GAMER in the non-relativistic limit, including the algorithm (\ref{app:cr_algorithm}) and numerical tests (\ref{app:cr_sound_wave} -- \ref{app:cr_ring}). See Section \ref{Methodology} for the simplification adopted in this paper for the relativistic case.

The governing equations of non-relativistic magnetohydrodynamics (MHD) with CR advection and anisotropic diffusion can be written as
\begin{subequations}
\begin{align}
&\frac{\partial \rho}{\partial t}+\nabla \cdot (\rho \boldsymbol{v})=0, \label{eq:CR_rho} \\
&\frac{\partial \rho \boldsymbol{v}}{\partial t}+\nabla \cdot \left( \rho \boldsymbol{v}\boldsymbol{v} - \frac{\boldsymbol{B}\boldsymbol{B}}{4\pi} \right) + \nabla p_{\text{total}}=0, \label{eq:CR_mom} \\
&\frac{\partial e}{\partial t}+\nabla \cdot \left[ (e+p_{\text{total}}) \boldsymbol{v} - \frac{\boldsymbol{B}(\boldsymbol{B} \cdot \boldsymbol{v})}{4\pi} \right] = \nabla \cdot (\boldsymbol{\kappa} \cdot \nabla e_{\text{cr}}), \label{eq:CR_energy} \\
&\frac{\partial e_{\text{cr}}}{\partial t}+\nabla \cdot (e_{\text{cr}}\boldsymbol{v}) = -p_{\text{cr}} \nabla \cdot \boldsymbol{v} + \nabla \cdot (\boldsymbol{\kappa} \cdot \nabla e_{\text{cr}}) \label{eq:CR_energy_cr}, \\
&\frac{\partial \boldsymbol{B}}{\partial t}-\nabla \times (\boldsymbol{v} \times  \boldsymbol{B})=0, \label{eq:CR_mag}
\end{align}
\end{subequations}
where $\rho$ is the gas mass density, $p_{\text{total}} = (\gamma_{\text{gas}} - 1)e_{\text{gas}} + (\gamma_{\text{cr}} - 1)e_{\text{cr}} + B^2 / 8\pi$ is the total pressure, $\boldsymbol{v}$ is the velocity, $\boldsymbol{B}$ is the magnetic field, $\boldsymbol{\kappa}$ is the diffusion tensor, and $e = 0.5\rho v^{2} + e_{\text{gas}} + e_{\text{cr}} + B^{2}/8\pi$ is the total energy density. Here $e_{\text{gas}}$ is the gas internal energy density, $\gamma_{\text{gas}}=5/3$ is the adiabatic index of an ideal gas, and $\gamma_{\text{cr}}=4/3$ is the effective adiabatic index of CRs in the relativistic limit \citep{CRFeedback_in_HD}. $p_{\text{cr}}=(\gamma_{\text{cr}} - 1)e_{\text{cr}}$ and $p_{\text{gas}}=(\gamma_{\text{gas}} - 1)e_{\text{gas}}$ are respectively the CR and gas pressure. The CR diffusion term can be written as
\begin{equation}
    \nabla \cdot (\boldsymbol{\kappa} \cdot \nabla e_{\text{cr}}) = \nabla \cdot (\kappa_{\parallel} \hat{\boldsymbol{b}} \hat{\boldsymbol{b}} \cdot \nabla e_{\text{cr}}) + \nabla \cdot [\kappa_{\bot} (\boldsymbol{I} - \hat{\boldsymbol{b}} \hat{\boldsymbol{b}}) \cdot \nabla e_{\text{cr}}],
\label{eq:CR_diffusion}
\end{equation}
where $\kappa_{\parallel}$ and $\kappa_{\bot}$ are the diffusion coefficients parallel and perpendicular to the magnetic field direction $\hat{\boldsymbol{b}}$, respectively \citep{1965RvPP....1..205B}, and $\boldsymbol{I}$ is the identity matrix.

\subsection{Algorithm}
\label{app:cr_algorithm}
The CR module adopts the VL scheme \citep{VL1, VL2} and the constraint transport technique \citep{Evans_1988} for solving Eqs. \ref{eq:CR_rho} -- \ref{eq:CR_mag}, supporting both piecewise linear and piecewise parabolic methods for data reconstruction \citep{1979JCoPh..32..101V, 1984JCoPh..54..174C}. The CR equation (Eq. \ref{eq:CR_energy_cr}) involves computing three terms: the advective flux $e_{\text{cr}}\boldsymbol{v}$, the adiabatic term $-p_{\text{cr}} \nabla \cdot \boldsymbol{v}$, and the anisotropic diffusion flux $\boldsymbol{\kappa} \cdot \nabla e_{\text{cr}}$. The advective flux is computed by regarding CR as a passive scalar advected along with the gas. The adiabatic term is computed as
\begin{align}
-p_{\text{cr}, i, j, k}^{n} \left[ \frac{v_{i+1/2, j, k}^{n} - v_{i-1/2, j, k}^{n}}{\Delta x} \right.
\left. + \frac{v_{i, j+1/2, k}^{n} - v_{i, j-1/2, k}^{n}}{\Delta y} + \frac{v_{i, j, k+1/2}^{n} - v_{i, j, k-1/2}^{n}}{\Delta z} \right], \label{eq:CR_work_half}
\end{align}
where $i, j, k$ are cell indices, $n$ is the discrete time step, and $\Delta x, \Delta y, \Delta z$ are the cell sizes. The cell-interface velocity, for example, $v_{i-1/2, j, k}^{n}$, is computed by an upwind method,
\begin{equation}
v_{i-1/2, j, k}^{n} = 
\left\{
  \begin{array}{lr}
    \frac{ F^{n, \rho}_{i-1/2, j, k} }{ \rho_{i-1, j, k} } & \text{if } F^{n, \rho}_{i-1/2, j, k} > 0 , \\
    \frac{ F^{n, \rho}_{i-1/2, j, k} }{ \rho_{i, j, k} } & \text{if } F^{n, \rho}_{i-1/2, j, k} \leq 0,
  \end{array}
\right.
\end{equation}
where $F^{n, \rho}_{i-1/2, j, k}$ is the cell-interface mass density flux between the cells $(i-1, j, k)$ and $(i, j, k)$.

We adopt a similar scheme as in \citet{Yang_2012} to compute the anisotropic diffusion flux in Eq. \ref{eq:CR_diffusion}. For example, the flux at the $i+1/2$ cell interface at time $n$ is given by
\begin{equation}
\label{eq:CR_diff}
F_{i+1/2, j, k}^{n, \text{diffusion}} = -\left( \kappa_{\parallel} - \kappa_{\bot} \right) b_{x} \left[ b_{x} \frac{\partial e_{\text{cr}}}{\partial x} + \overline{b_{y}} \overline{\frac{\partial e_{\text{cr}}}{\partial y}} + \overline{b_{z}} \overline{\frac{\partial e_{\text{cr}}}{\partial z}} \right] -  \kappa_{\bot} \frac{\partial e_{\text{cr}}}{\partial x},
\end{equation}
\begin{align}
b_{x} &= b_{x, i+1/2, j, k}^{n}, \label{eq:diff_magx} \\
\overline{b_{y}} &= ( b_{y, i, j-1/2, k}^{n} + b_{y, i+1, j-1/2, k}^{n} + b_{y, i, j+1/2, k}^{n} + b_{y, i+1, j+1/2, k}^{n}) /4, \label{eq:diff_magy} \\
\overline{b_{z}} &= ( b_{z, i, j, k-1/2}^{n} + b_{z, i+1, j, k-1/2}^{n} + b_{z, i, j, k+1/2}^{n} + b_{z, i+1, j, k+1/2}^{n}) /4, \label{eq:diff_magz}
\end{align}
\begin{equation}
\label{eq:diff_slope_x}
\frac{\partial e_{\text{cr}}}{\partial x} = \frac{ e_{\text{cr}, i+1, j, k} - e_{cr, i, j, k} }{\Delta x},
\end{equation}
\begin{align}
\label{eq:diff_slope_y}
\overline{\frac{\partial e_{\text{cr}}}{\partial y}} = 
        L  \left[ 
            L\left( \left. \frac{\partial e_{\text{cr}}}{\partial y} \right|_{i, j-1/2, k}, 
                    \left. \frac{\partial e_{\text{cr}}}{\partial y} \right|_{i, j+1/2, k} \right), \right.
           \left. L\left( \left. \frac{\partial e_{\text{cr}}}{\partial y} \right|_{i+1, j-1/2, k},
                    \left. \frac{\partial e_{\text{cr}}}{\partial y} \right|_{i+1, j+1/2, k} \right)
         \right],
\end{align}
\begin{align}
\label{eq:diff_slope_z}
\overline{\frac{\partial e_{\text{cr}}}{\partial z}} = 
         L \left[ 
             L\left( \left. \frac{\partial e_{\text{cr}}}{\partial z} \right|_{i, j, k-1/2},
                     \left. \frac{\partial e_{\text{cr}}}{\partial z} \right|_{i, j, k+1/2} \right), \right .
      \left. L\left( \left. \frac{\partial e_{\text{cr}}}{\partial z} \right|_{i+1, j, k-1/2},
                     \left. \frac{\partial e_{\text{cr}}}{\partial z} \right|_{i+1, j, k+1/2} \right)
         \right],
\end{align}
where the diffusion coefficient $\boldsymbol{\kappa}$ is taken to be a constant. $L$ is a slope limiter, for which we adopt the monotonized central limiter,
\begin{equation} \label{eq:MC_limiter}
L(a, b) = \text{minmod} \left[ 2 \text{minmod}(a, b), \frac{a+b}{2} \right],
\end{equation}
\begin{equation}
\text{minmod}(a, b) = 
\left\{
  \begin{array}{lr}
    \text{min}(a, b) & \text{if } a, b > 0, \\
    \text{max}(a, b) & \text{if } a, b < 0, \\
    0 & \text{if } ab \leq 0.
  \end{array}
\right.
\end{equation}

We add the CR diffusion flux to the CR advective flux and the total energy flux when updating $e_{\text{cr}}$ and $e$ in Eqs. \ref{eq:CR_energy} and \ref{eq:CR_energy_cr} in a conservative fashion. All CR terms are incorporated into both the half- and full-time-step updates in the VL scheme, which is necessary to preserve second-order accuracy (see Section \ref{app:cr_sound_wave}).

The simulation time-step $\Delta t$ must satisfy the following Courant–Friedrichs–Lewy (CFL) condition,
\begin{equation}
\Delta t \leq C_{\text{CFL}}\frac{\text{min} \left[ \Delta x^{2}, \Delta y^{2}, \Delta z^{2} \right] }{\text{max}\left( \kappa_{\parallel }, \kappa_{\bot } \right)},
\end{equation}
where $C_{\text{CFL}}$ is the safety factor with a default value of $0.3$.

By taking advantage of the flexibility and extensibility of GAMER, the CR module directly inherits the AMR structure and the hybrid MPI/OpenMP/GPU parallelization framework from the previous non-relativistic MHD and special-relativistic hydrodynamics implementations \citep{Zhang_2018, tseng2021}. Especially, the CPU/GPU integration infrastructure in GAMER allows the same source codes to be used for CPU-only and GPU-accelerated computations, avoiding redundant code development and simplifying code maintenance.

\subsection{Linear Sound Wave Test}
\label{app:cr_sound_wave}
We use a linear sound wave to validate the coupling between CRs and gas and to measure the error convergence rate \citep{Rasera_2008}.
Small perturbations $\delta \rho$, $\delta p_{\text{gas}}$, $\delta p_{\text{cr}}$, and $\delta v$ satisfying
\begin{equation}
\frac{\delta \rho}{\rho_{0}} = \frac{\delta p_{\text{gas}}}{\gamma_{\text{gas}} p_{\text{gas}, 0}}  = \frac{\delta p_{\text{cr}}}{\gamma_{cr} p_{cr, 0}} = \frac{\delta v}{c_{s}}
\end{equation}
are added to a homogeneous background $\rho_0=1$, $p_{\text{gas},0}=1$, $p_{\text{cr},0}=1$, and $v_0=1$,
with $\delta \rho/\rho_0=10^{-6}$ to avoid any non-linear effect.
Neither magnetic field nor CR diffusion is included in this test.
The effective sound speed $c_{s}$ is given by
\begin{equation}
c_{s}=\sqrt{ \frac{\gamma_{\text{gas}} p_{\text{gas}, 0} + \gamma_{\text{cr}} p_{\text{cr},0}}{\rho_0} }.
\end{equation}
The periodic computational domain is $\{x, y, z\} \in [0, 1]$ with a uniform resolution of $N = 32 \text{--} 512$ cells along each dimension.
The sound wave travels along the diagonal direction with a wavelength of $1/\sqrt{3}$.

\begin{figure}
\centering
\includegraphics[width=\textwidth]{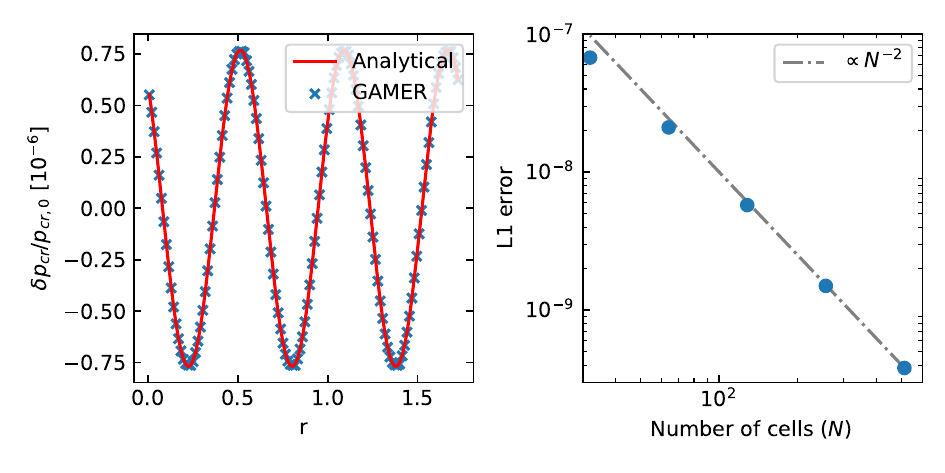}
\caption{CR linear sound wave test. The left panel shows the CR pressure perturbation in the diagonal direction after evolving for one period with $N=128$. Crosses are the simulation data and the solid line shows the analytical solution. The right panel shows the L1 error, demonstrating second-order accuracy.}
\label{fig:L1error_FLASH_GAMER_3D_withV_final}
\end{figure}

Fig. \ref{fig:L1error_FLASH_GAMER_3D_withV_final} compares the simulation results and the analytical solution with $N=128$ after evolving for one period (left panel) and shows the L1-norm error as a function of $N$ (right panel). Simulation data agree well with the analytical solution and achieve second-order accuracy.

\subsection{Shock Tube Test}
\label{app:cr_shocktube}
The shock tube problem provides a standard test for nonlinear evolution and shock capturing in computational hydrodynamics. For a hybrid fluid comprising gas and CRs, an analytical solution has been derived by \citet{10.1111/j.1365-2966.2005.09953.x}. We perform one-dimensional simulations in a computational domain $x \in [0, 1]$ with outflow boundary conditions and a uniform resolution of $N=1024$ cells. The left state of the fluid variables $(0<x<0.5)$ is initialized as $\rho_{L} = 1.0$, $\boldsymbol{v}_L = 0.0$, $p_{\text{gas}, L} = 6.7 \times 10^4$, and $p_{\text{cr}, L} = 1.3 \times 10^5$; the right state $(0.5 \leq x<1)$ is initialized as $\rho_{R} = 0.2$, $\boldsymbol{v}_R = 0.0$, $p_{\text{gas}, R} = 2.4 \times 10^2$, and $p_{\text{cr}, R} = 2.4 \times 10^2 $. Neither magnetic field nor CR diffusion is included in this test.

\begin{figure}
\centering
\includegraphics[width=\textwidth]{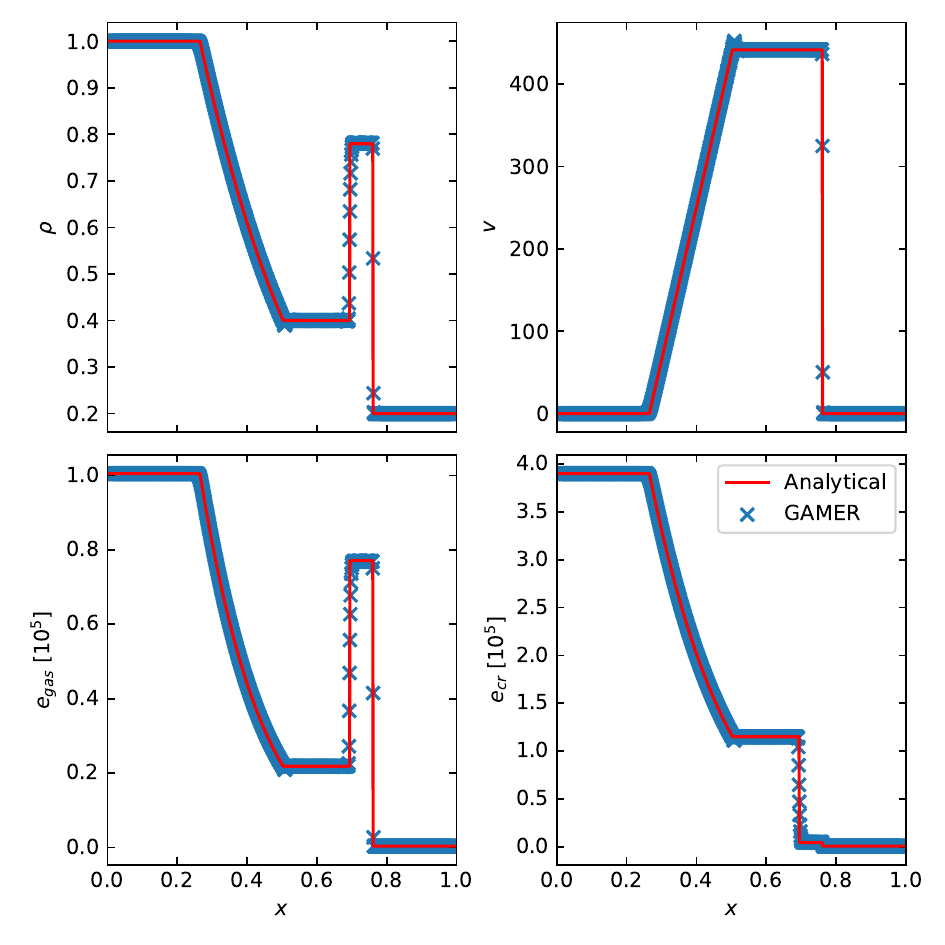}
\caption{CR shock tube test. It compares the gas mass density (top left), velocity (top right), gas internal energy density (bottom left), and CR energy density (bottom right) between the simulation results (crosses) and the analytical solution (solid lines).}
\label{fig:Shocktube}
\end{figure}

Fig. \ref{fig:Shocktube} shows the simulation results and the analytical solution at $t=4.4 \times 10^{-4}$. The simulation results match well with the analytical solution, including the rarefaction wave, the contact discontinuity, and the shock.

\subsection{Ring Test for Anisotropic CR Diffusion}
\label{app:cr_ring}
The ring test simulates the anisotropic CR diffusion along a circular magnetic field \citep{Parrish_2005, Sharma_2007, Yang_2012, Jiang_2018}. To simulate pure CR diffusion, all the fluid variables except for $e_{\text{cr}}$ are fixed, $\boldsymbol{v}=0.0$, and the adiabatic term $-p_{\text{cr}} \nabla \cdot \boldsymbol{v}$ is ignored. The magnetic field is set to $\boldsymbol{B}=(-y/r, x/r, 0)$ with $r=\sqrt{x^2+y^2}$. The CR energy density is initialized as
\begin{equation}
\label{eq:ring_cr_ini}
    e_{\text{cr}}(x, y) = 
    \left\{
      \begin{array}{lr}
        12 & \text{ if } 0.5<r<0.7 \text{ and } |\phi| \leq \frac{\pi}{12}, \\
        10 & \text{ otherwise,}
      \end{array}
    \right.
\end{equation}
where $\phi=\text{atan2}(y,x)$ is the azimuthal angle. The diffusion coefficients are $\kappa_{\parallel} = 0.05$ and $\kappa_{\bot} = 0.0$.
The analytical solution at time $t$ is given by
\begin{equation}
\label{eq:ring_ana}
    e_{\text{cr}}(x, y, t) =
    \left\{
      \begin{array}{lr}
        10 + \text{Erfc}\left[\left(\phi-\frac{\pi}{12}\right)\frac{r}{\sqrt{4t\kappa_{\parallel}}}\right]
           - \text{Erfc}\left[\left(\phi+\frac{\pi}{12}\right)\frac{r}{\sqrt{4t\kappa_{\parallel}}}\right] & \text{ if } 0.5<r<0.7, \\
        10 & \text{ otherwise,}
      \end{array}
    \right.
\end{equation}
where $\text{Erfc}(x)$ is the complementary error function. The computation domain is $\{x, y\} \in [-1, 1]$ with a uniform resolution of $256^2$ cells.

\begin{figure}
\centering
\includegraphics[width=\textwidth]{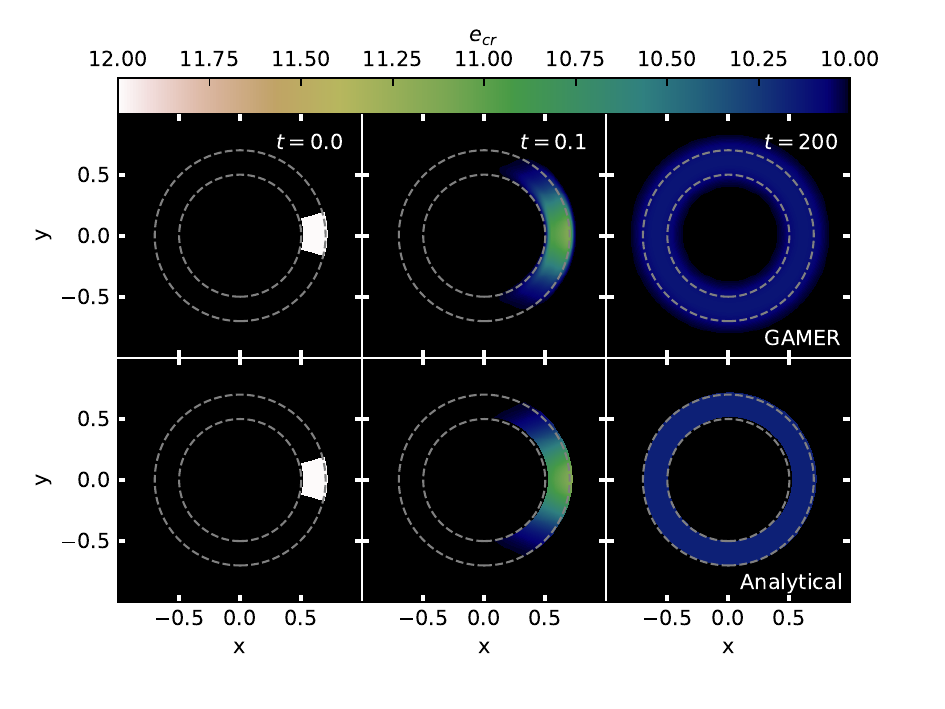}
\caption{CR ring test for anisotropic diffusion along a circular magnetic field. The top and bottom rows show the simulation results and the analytical solutions of CR energy density, respectively. Columns from left to right correspond to different simulation times: $t=0$, $0.1$, and $200$. The dashed circles with $r=0.5$ and $0.7$ enclose the region where diffusion occurs.}
\label{fig:Ring_Test_time_series}
\end{figure}

Fig. \ref{fig:Ring_Test_time_series} plots the simulation results at $t=0$, $0.1$, and $200$, showing good agreement with the analytical solutions. For the long-term solution at $t=200$, diffusion along the azimuthal direction results in a homogeneous distribution of CRs within the ring, despite a small amount of numerical diffusion perpendicular to the magnetic field even with $\kappa_{\bot} = 0.0$ due to Cartesian grids.

\bibliography{sample631}{}
\bibliographystyle{aasjournal}

\end{document}